\documentclass[reqno]{amsart}
\usepackage{amssymb}

\newtheorem{theorem}{Theorem}[section]
\newtheorem{proposition}[theorem]{Proposition}
\newtheorem{lemma}[theorem]{Lemma}
\newtheorem{corollary}[theorem]{Corollary}

\theoremstyle{definition}

\theoremstyle{remark}

\newtheorem*{note}{Note}

\numberwithin{equation}{section}

\begin{document}

\title[Scattering on a Weyl Chamber]
{Scattering Theory of Discrete (Pseudo) Laplacians on a Weyl
Chamber}

\author{J.F. van Diejen}
\address{
Instituto de Matem\'atica y F\'{\i}sica, Universidad de Talca,
Casilla 747, Talca, Chile}

\thanks{Work supported in part by the Fondo Nacional de Desarrollo
Cient\'{\i}fico y Tecnol\'ogico (FONDECYT) Grant \# 1010217 and
the Programa Formas Cuadr\'aticas of the Universidad de Talca.}


\begin{abstract}
To a crystallographic root system we associate a system of
multivariate orthogonal polynomials diagonalizing an integrable
system of discrete pseudo Laplacians on the Weyl chamber. We
develop the time-dependent scattering theory for these discrete
pseudo Laplacians and determine the corresponding wave operators
and scattering operators in closed form. As an application, we
describe the scattering behavior of certain hyperbolic
Ruijsenaars-Schneider type lattice Calogero-Moser models
associated with the Macdonald polynomials.
\end{abstract}

\maketitle \vspace{-1ex}
\tableofcontents

\section{Introduction}\label{sec1}
A fundamental property of the solitonic solutions of integrable
nonlinear wave equations is that their multi-particle scattering
process decomposes into pairwise two-particle interactions
\cite{sco-chu-mcl:soliton,abl-seg:solitons,nmpz:theory,new:solitons,%
fad-tak:hamiltonian}.
This phenomenon is preserved at the quantum level: the
corresponding solitonic quantum field theories are characterized
by an $N$-particle scattering matrix that factorizes in terms of
two-particle scattering matrices
\cite{mat-lie:many-body,kor-bog-ize:quantum}. As it turns out,
this type of factorization can be understood heuristically as
being a consequence of the integrability of the models in question
\cite{kul:factorization,rui-sch:new}.

An archetype example of an integrable system with factorized
scattering is the celebrated nonlinear Schr\"odinger equation
(NLS). The quantum version of this model boils down to a bosonic
$N$-particle system with a pairwise interaction via
delta-functional potentials. The factorization of the scattering
manifests itself through the asymptotics of the wave function,
which is characterized by (products of) two-particle scattering
matrices (or $c$-functions)
\cite{mat-lie:many-body,gau:fonction,oxf:hamiltonian,%
kor-bog-ize:quantum}.

In recent work, Ruijsenaars constructed a remarkably large class
of quantum integrable lattice models of $N$-particles exhibiting
factorized scattering \cite{rui:factorized}. The discrete systems
in question arise by interpreting recurrence relations (or Pieri
formulas) for symmetric multivariate orthogonal polynomials as
quantum eigenvalue equations. Here the polynomial variable plays
the role of the spectral parameter and the index (i.e. partition)
labelling the polynomials is thought of as the discrete spatial
variable. By analyzing the asymptotics of the polynomials as the
degree tends to infinity, Ruijsenaars demonstrated that---for
factorized orthogonality measures subject to certain technical
conditions ensuring that the particle interaction is short-range
and the spectrum is absolutely continuous---the corresponding
discrete models are governed by a scattering matrix that
factorizes into two-particle scattering matrices. An interesting
particular case is that of the Macdonald polynomials
\cite{mac:symmetric}. The corresponding $N$-particle model can  be
identified as a hyperbolic Ruijsenaars-Schneider type lattice
Calogero-Moser system \cite{rui:finite-dimensional,rui:systems}.
At the level of classical Hamiltonian mechanics, the scattering of
the corresponding integrable system was studied in great detail in
Ref. \cite{rui:action-angle}.

It is known that the root systems of simple Lie algebras form a
fruitful context for understanding Calogero-Moser systems and
particle models with delta-functional potentials
\cite{ols-per:quantum,gut:integrable,hec-sch:harmonic,%
hec-opd:yang,opd:lecture}.
From this perspective, it is natural to ask for a generalization
of Ruijsenaars' construction to the case of arbitrary root
systems. The purpose of the present paper is to provide such a
construction.

More specifically, we associate to a
crystallographic root system a system of Weyl-group invariant
multivariate orthogonal polynomials on the Weyl alcove,
characterized by a weight function that factorizes over the roots
(of the root system) in terms of one-dimensional $c$-functions.
The orthogonality implies that the polynomials satisfy a system of
recurrence relations (Pieri formulas). These
recurrence relations are interpreted
as eigenvalue equations for an integrable
system of discrete pseudo Laplacians on the Weyl chamber. We
develop the time-dependent scattering theory for these discrete
pseudo Laplacians and determine the corresponding wave operators
and scattering operators in closed form.
For a specific choice of
the weight function, our polynomials amount to the Macdonald
polynomials associated with root systems
\cite{mac:orthogonal,mac:affine}. Again the
corresponding integrable lattice model then permits identification
as a discrete hyperbolic Ruijsenaars-Schneider type Calogero-Moser
system
\cite{rui:finite-dimensional,rui:systems,die:integrability}. For
the type $A$ root systems the Weyl group is the symmetric group and
we reproduce the results of Ruijsenaars \cite{rui:factorized}.

The wave- and scattering operators computed in this paper
compare the dynamics generated
by the discrete pseudo Laplacian to that of a free discrete
Laplacian (corresponding to
the case that the $c$-functions reduce to constant functions).
Our study of the scattering consists of two parts. In the first
(time-independent)
part it is shown that the wave function of the discrete
pseudo Laplacian has plane wave asymptotics, provided that the $c$-functions
determining the orthogonality measure of the polynomials
satisy certain analyticity
requirements (guaranteeing that the spectrum of the discrete pseudo
Laplacian is absolutely continuous). This part of the
discussion hinges on previous results describing the large-degree
asymptotics of the class of multivariate orthogonal polynomials
under consideration \cite{die:asymptotic,die:asymptotics}.
The second (time-dependent)
part consists of a stationary phase analysis that permits proving
the existence and unitarity of the wave operators and scattering
operators given the plane wave asymptotics of the wave functions.
Key ingredient of this part of the discussion is a stationary phase estimate
from \cite[p. 38-39]{ree-sim:methods} that controls the decay for
$t\to\pm\infty$ of certain
oscillatory integrals describing the difference between interacting and freely
evolving wave packets.

The paper is organized as follows. Section \ref{sec2} describes
the construction of orthogonal polynomials related to root
systems. In Section \ref{sec3} we introduce a commuting system of
discrete pseudo Laplacians on the Weyl chamber diagonalized by the
orthogonal polynomials in question. The wave operators and
scattering operators for our discrete pseudo Laplacians are
determined in Section \ref{sec4}.  The stationary phase analysis
that lies at the basis of the computation of these wave-- and
scattering operators is relegated to Section \ref{sec5}. Finally,
in Section \ref{sec6} we specialize to the case of Macdonald
polynomials and detail the scattering theory of the associated
hyperbolic Ruijsenaars-Schneider type lattice Calogero-Moser
models. Some key properties of the Macdonald polynomials invoked
in Section \ref{sec6} have been collected in Appendix \ref{appA}
at the end of the paper. For the reader's convenience, we have
also included an index of notations in Appendix \ref{appB}.

Let us conclude this introduction by providing a brief description
of what the main results amount to in the elementary (classical)
situation of a root system of
rank 1. Let $\hat{c}(z)$ be a zero-free analytic function on the
disc $|z|\leq \varrho$, with $\varrho >1$, that is real-valued for $z$
real and normalized such that $\hat{c}(0)=1$. We associate to
$\hat{c}(z)$ an orthonormal basis of trigonometric polynomials
$P_0(\xi), P_1(\xi),P_2(\xi),\ldots$ for the Hilbert space
$L^2((0,\pi),\frac{2\sin^2(\xi)\text{d}x}
{\pi\hat{c}(e^{i\xi})\hat{c}(e^{-i\xi})})$ that is obtained by
applying the Gram-Schmidt process to the Fourier-cosine basis
$1,\cos (\xi),\cos (2\xi),\ldots$. It is an immediate consequence
of the three-term recurrence relation for the orthonormal
polynomials $P_\ell (\xi)$ that the wave function
\begin{equation}\label{wave}
\Psi_\ell (\xi)=
\frac{2\sin(\xi)P_\ell (\xi)}{\sqrt{\hat{c}(e^{i\xi})\hat{c}(e^{-i\xi})}},
\quad \xi\in (0,\pi),\; \ell\in\mathbb{N},
\end{equation}
satisfies an eigenvalue equation of the form
$L\Psi=2\cos(\xi)\Psi$, where $L$ represents a discrete
(self-adjoint) Laplacian acting on lattice functions $\phi\in
\ell^2 (\mathbb{N})$ as
\begin{equation}\label{lap}
L\phi_\ell = a_\ell
\phi_{\ell+1}+b_\ell\phi_\ell+a_{\ell-1}\phi_{\ell-1} \qquad
(\phi_{-1}\equiv 0),
\end{equation}
with $a_\ell, b_\ell $ denoting the coefficients of the three-term
recurrence relation. For $\hat{c}(z)=1$, the polynomials
$P_\ell(\xi)$ amount to the Chebyshev polynomials of the second
kind $U_\ell (\cos \xi)=\sin (\ell+1)\xi/\sin\xi$, whence the wave
function in Eq. \eqref{wave} reduces in this case to the
Fourier-sine kernel $\Psi_\ell^{(0)}(\xi)=2\sin (\ell +1)\xi$. The
Laplacian $L$ \eqref{lap} then amounts to a free Laplacian
$L^{(0)}$ whose action on lattice functions is given by
$L^{(0)}\phi_\ell=\phi_{\ell +1}+\phi_{\ell-1}$.

Theorem \ref{plane:thm} (below) now states that  for $\ell\to\infty
$ the wave function $\Psi_\ell (\xi)$ \eqref{wave} converges
exponentially fast in $L^2((0,\pi),(2\pi)^{-1}\text{d}\xi )$ to
the anti-symmetric combination of plane waves
\begin{equation}\label{pw-exp}
\Psi_l^{\infty}(\xi)=\hat{s}^{1/2}(\xi) e^{i(\ell+1)\xi}-\hat{s}^{-1/2}(\xi)
e^{-i(\ell+1)\xi},
\end{equation}
with
$\hat{s}(\xi )=\hat{c}(e^{-i\xi})/\hat{c}(e^{i\xi} )$.

Furthermore, let us denote by $\mathcal{F}:l^2(\mathbb{N})\mapsto
L^2((0,\pi),(2\pi)^{-1}\text{d}\xi)$ and
$\mathcal{F}^{(0)}:l^2(\mathbb{N})\mapsto
L^2((0,\pi),(2\pi)^{-1}\text{d}\xi)$ the Fourier pairings with
kernel $\Psi_\ell(\xi)$ and $\Psi_\ell^{(0)}(\xi)$, respectively:
\begin{equation}
\begin{cases}
{\displaystyle \hat{\phi}(\xi) = \sum_{\ell\in\mathbb{N}}
\phi_\ell\Psi_\ell(\xi)}  \\[2ex]
{\displaystyle \phi_\ell = \frac{1}{2\pi} \int_0^\pi
\hat{\phi}(\xi)\Psi_\ell(\xi)\text{d}\xi }
\end{cases} ,
\qquad
\begin{cases}
{\displaystyle \hat{\phi}(\xi) = \sum_{\ell\in\mathbb{N}}
\phi_\ell\Psi^{(0)}_\ell(\xi)}  \\[2ex]
{\displaystyle \phi_\ell = \frac{1}{2\pi} \int_0^\pi
\hat{\phi}(\xi)\Psi_\ell^{(0)}(\xi)\text{d}\xi }
\end{cases}.
\end{equation}
Then Theorem \ref{wave:thm} and Corollary \ref{scattering:cor}
(below) state that the wave operators $\Omega_\pm=s-\lim_{t\to
\pm\infty} e^{itL} e^{-itL^{(0)}}$ and the scattering operator
$\mathcal{S}=\Omega_+^{-1}\Omega_-$ exist in $l^2(\mathbb{N})$ and
are given explicitly by the unitary operators $\Omega_\pm =
\mathcal{F}^{-1}\circ \hat{\mathcal{S}}^{\mp 1/2}\circ
\mathcal{F}^{(0)}$ and $\mathcal{S} =
(\mathcal{F}^{(0)})^{-1}\circ \hat{\mathcal{S}}\circ
\mathcal{F}^{(0)}$, where $\hat{\mathcal{S}}$ denotes a unitary
scattering matrix that is characterized by its multiplicative action on a
wave packet $\hat{\phi}\in L^2((0,\pi),(2\pi)^{-1}\text{d}\xi)$ of
the form $(\hat{\mathcal{S}} \hat{\phi})(\xi)=
\hat{s}(-\xi)\hat{\phi}(\xi) $ for $0 <\xi <\pi$ (with $\hat{s}(\xi)$ as defined
just below Eq. \eqref{pw-exp}).

\vspace{3ex} {\bf Acknowledgments.} Thanks are due to S.N.M.
Ruijsenaars for several helpful discussions and to the referees
for suggesting some improvements in the presentation.

\section{Orthogonal Polynomials Related to Root Systems}\label{sec2}
In this section we introduce a class of multivariate orthogonal
polynomials related to root systems. For basic facts on root
systems we refer to the standard works
\cite{bou:groupes,hum:introduction}.

\subsection{Polynomials on the Weyl Alcove}\label{sec2.1}
Let $\mathbf{E}$, $\langle\cdot ,\cdot\rangle$ be a real
$N$-dimensional Euclidean vector space and let
$\boldsymbol{R}\subset \mathbf{E}$ denote an irreducible
crystallographic root system spanning $\mathbf{E}$. We write
$\mathcal{Q}$ and $\mathcal{Q}^+$ for the root lattice and its
nonnegative semigroup generated by the positive roots
$\boldsymbol{R}^+$
\begin{equation}
\mathcal{Q}= \text{Span}_\mathbb{Z}
(\boldsymbol{R}),\;\;\;\mathcal{Q}^+= \text{Span}_\mathbb{N}
(\boldsymbol{R}^+) ,
\end{equation}
and we write $\mathcal{P}$ and $\mathcal{P}^+$ for the weight
lattice its nonnegative cone of dominant weights
\begin{subequations}
\begin{eqnarray}
\mathcal{P} &=& \{ \lambda\in \mathbf{E} \mid  \langle \lambda
,\alpha^\vee
\rangle \in\mathbb{Z},\; \forall \alpha\in \boldsymbol{R} \} ,\\
 \mathcal{P}^+ &=&
\{ \lambda\in \mathbf{E} \mid  \langle \lambda ,\alpha^\vee
\rangle \in\mathbb{N},\; \forall \alpha\in \boldsymbol{R}^+ \} ,
\end{eqnarray}
\end{subequations}
where we have introduced the coroot $\alpha^\vee\equiv 2\alpha
/\langle \alpha ,\alpha \rangle$. The algebra of (trigonometric)
polynomials on the Weyl alcove
\begin{equation}
\mathbf{A}= \{ \xi\in \mathbf{E} \mid 0 <\langle \xi ,\alpha
\rangle < 2\pi, \; \forall \alpha\in \boldsymbol{R}^+ \}
\end{equation}
is spanned by the basis of the monomial symmetric functions
\begin{equation}
m_\lambda (\xi) = \frac{1}{| W_\lambda |} \sum_{w\in W} e^{i
\langle \lambda , \xi_w \rangle },\qquad \lambda\in \mathcal{P}^+,
\end{equation}
where $W\subset \text{GL}(\mathbf{E})$ denotes the Weyl group of
the root system $\boldsymbol{R}$, $\xi_w\equiv w(\xi)$, and
$|W_\lambda|$ stands for the order of the stabilizer subgroup
$W_\lambda =\{ w\in W \mid w(\lambda) =\lambda \}$.

\subsection{Factorized Weight Functions}\label{sec2.2} We will now introduce a
class of smooth weight functions on the Weyl alcove $\mathbf{A}$
that factorize over the root system $\boldsymbol{R}$. To this end
we write $\boldsymbol{R}_0=\{ \alpha \in \boldsymbol{R} \mid
2\alpha \not\in \boldsymbol{R}\}$ and $\boldsymbol{R}_1=\{ \alpha
\in \boldsymbol{R} \mid \frac{\alpha}{2} \not\in
\boldsymbol{R}\}$. (So for a reduced root system one has that
$\boldsymbol{R}_0=\boldsymbol{R}_1=\boldsymbol{R}$ and for the
nonreduced root system $\boldsymbol{R}=BC_N$ one has that
$\boldsymbol{R}_0=C_N$ and $\boldsymbol{R}_1=B_N$.) The weight
functions under consideration are of the form
\begin{subequations}
\begin{equation}\label{p-m}
\hat{\Delta} (\xi)=\frac{1}{\hat{\mathcal{C}} (\xi)
\hat{\mathcal{C}}(-\xi)},
\end{equation}
with
\begin{equation}\label{c-f}
\hat{\mathcal{C}} (\xi) =
 \prod_{\alpha \in
\boldsymbol{R}_1^+} \hat{c}_{|\alpha|}(e^{-i\langle \alpha ,\xi
\rangle} ) ,
\end{equation}
\end{subequations}
where it assumed that the $c$-functions $\hat{c}_{|\alpha|}(z)$
building $\hat{\mathcal{C}}(\xi)$ \eqref{c-f} depend only on the
length of the root $\alpha$ (so
$\hat{c}_{|\alpha|}(z)=\hat{c}_{|\beta|}(z)$ if $\alpha$ and
$\beta$ lie on the same Weyl-orbit). For technical reasons, we
will furthermore assume that these $c$-functions $
\hat{c}_{|\alpha|}(z)$ are {\em (i)} analytic and zero-free on a
closed disc $\mathbb{D}_\varrho =\{ z\in\mathbb{C}\mid |z|\leq
\varrho \}$ of radius $\varrho
>1$, {\em (ii)} normalized such that $ \hat{c}_{|\alpha|}(0)=1$,
and  {\em (iii )} real-valued for $z\in\mathbb{R}$ (so
$\hat{\mathcal{C}}(-\xi)=\overline{\hat{\mathcal{C}}(\xi)}$).

\subsection{Gram-Schmidt Orthogonalization}\label{sec2.3}
The technical conditions on the $c$-func\-tions ensure that
$\hat{\Delta} (\xi)$ \eqref{p-m}, \eqref{c-f} defines a smooth
positive weight function on $\mathbf{A}$ (which extends
analytically to a Weyl-group invariant function on $\mathbf{E}$).
We employ this weight function to endow the space of trigonometric
polynomials on the Weyl alcove with an inner product structure via
embedding in the Hilbert space
$L^2(\mathbf{A},\hat{\Delta}|\delta|^2\, \text{d}\xi)$:
\begin{equation}
( f , g)_{\hat{\Delta}} =
 \frac{1}{|W|\, \text{Vol}(\mathbf{A})  }\,
\int_{\mathbf{A}} f(\xi) \overline{g(\xi)}\,\hat{\Delta}
(\xi)\,|\delta(\xi) |^2 \text{d} \xi ,\qquad \forall f,g\in
L^2(\mathbf{A},\hat{\Delta}|\delta|^2\,\text{d}\xi),
\end{equation}
where $\overline{g(\xi)}$ stands for the complex conjugate of
$g(\xi)$, $\text{Vol}(\mathbf{A})=\int_{\mathbf{A}}\text{d}\xi$,
and $\delta (\xi)$ denotes the Weyl denominator
\begin{equation} \label{delta} \delta(\xi)= \prod_{\alpha \in \boldsymbol{R}_0^+} (e^{i
\langle \alpha ,\xi \rangle /2}-e^{-i \langle \alpha ,\xi \rangle
/2}) .
\end{equation}

Let $\succeq$ be a (partial) order of the dominant weights
$\mathcal{P}^+$ refining the {\em dominance partial order}
\begin{equation}\label{po}
\lambda \geqslant \mu \;\Longleftrightarrow \;\lambda -\mu \in
\mathcal{Q}^+
\end{equation}
such that the highest-weight spaces $\text{Span} \{ m_\mu
\}_{\mu\in\mathcal{P}^+,\mu\preceq\lambda}$ remain
finite-dimensional for all $\lambda \in\mathcal{P}^+$. By applying
the Gram-Schmidt process to the partially ordered monomial basis
$\{ m_\lambda \}_{\lambda\in \mathcal{P}^+}$, we construct a
normalized basis $\{ P_\lambda \}_{\lambda\in \mathcal{P}^+}$ of
$L^2(\mathbf{A},\hat{\Delta}|\delta|^2\,\text{d}\xi)$ given by
trigonometric polynomials of the form
\begin{subequations}
\begin{equation}\label{op1}
P_\lambda (\xi) =  \sum_{\mu\in\mathcal{P}^+,\, \mu \preceq
\lambda} a_{\lambda\mu} m_\mu (\xi),\qquad
\lambda\in\mathcal{P}^+,
\end{equation}
with coefficients $a_{\lambda\mu}\in\mathbb{C}$ such that
\begin{equation}\label{op2}
( P_\lambda , P_\mu )_{\hat{\Delta}} =
\begin{cases}
0  & \text{if}\; \mu \prec\lambda , \\
1 &\text{if}\; \mu =\lambda
\end{cases}
\end{equation}
\end{subequations}
(where $ a_{\lambda \lambda}>0$ by convention). The Gram-Schmidt
process guarantees that the polynomials $P_\lambda$,
$\lambda\in\mathcal{P}^+$ are orthogonal when comparable in the
(partial) order $\succeq$ (i.e.  $( P_\lambda , P_\mu
)_{\hat{\Delta}} = 0$ when $\lambda \succ \mu$ or
$\lambda\prec\mu$). Hence, a sufficient condition to ensure that
our polynomials form an orthonormal basis of the Hilbert space
$L^2(\mathbf{A}, \hat{\Delta} |\delta|^2\,\text{d}x)$ is to
require the refinement $\succeq$ of the dominance order
$\geqslant$ to be a {\em linear} ordering of $\mathcal{P}^+$.
 (The fact that the
polynomials in Eqs. \eqref{op1}, \eqref{op2}
form a complete set in $L^2(\mathbf{A}, \hat{\Delta} |\delta|^2\,\text{d}x)$
is a consequence of the
Stone-Weierstrass theorem.) In
general, the orthonormal basis in question depends on the choice
of such linear refinement. It will turn out below, however, that
for our principal applications the $c$-functions are such that the
orthogonality is already guaranteed when taking for $\succeq$
simply the dominance ordering $\geqslant$ \eqref{po} itself (in
other words, in such case the construction results to be
independent of the choice of the linear refinement). From now on
we will always assume that we have fixed a sufficiently fine
(partial) ordering $\succeq$ so as to guarantee that the basis $\{
P_\lambda \}_{\lambda\in \mathcal{P}^+}$ be {\em orthogonal} (i.e.
$( P_\lambda , P_\mu )_{\hat{\Delta}} = 0$ when $\lambda \neq
\mu$).

\subsection{Weyl Characters}\label{weyl:sec}\label{sec2.4}
The simplest example of the above construction is the special case
with unit $c$-functions, i.e., with $\hat{c}_{|\alpha|} (z)=1$,
$\forall \alpha \in \boldsymbol{R}_1^+$. The weight function then
becomes of the form $\hat{\Delta} (\xi)=1$ and the Gram-Schmidt
process turns out to be independent of the choice of the
refinement $\succeq$ of $\geqslant$ (i.e. in this case we may take
$\succeq$ to be equal to $\geqslant$ without restriction). The
corresponding orthonormal polynomials $P_\lambda (\xi)$ amount to
the celebrated Weyl characters \cite{mac:orthogonal,mac:affine}
\begin{equation}\label{weylcars}
P_\lambda (\xi)= \chi_\lambda (\xi)\equiv \delta^{-1} (\xi)
\sum_{w\in W} (-1)^w\, e^{i\langle \rho+\lambda , \xi_w\rangle }
,\qquad \lambda\in\mathcal{P}^+,
\end{equation}
where $(-1)^w\equiv\det (w)$ and $\rho
\equiv\frac{1}{2}\sum_{\alpha\in \boldsymbol{R}_0^+}\alpha$.

For later use, it will actually be convenient to extend the
definition of the Weyl characters $\chi_\lambda (\xi )$ in Eq.
\eqref{weylcars} to the case of nondominant weights $\lambda$. It
is immediate from this definition that for
$\lambda\in\mathcal{P}\setminus\mathcal{P}^+$
\begin{equation}\label{weylcars2}
\chi_\lambda (\xi ) =
\begin{cases}
(-1)^{w_{\rho+\lambda}} \chi_{w_{\rho+\lambda}
(\rho+\lambda)-\rho}(\xi)
&\text{if}\;\; |W_{\rho+\lambda}| =1,  \\
0& \text{if}\;\; |W_{\rho+\lambda}| >1,
\end{cases}
\end{equation}
where, for $\mu\in\mathcal{P}$ regular,  $w_\mu\in W$ denotes the
unique Weyl group element such that $w_\mu (\mu
)\in\mathcal{P}^+$.

\section{Discrete (Pseudo) Laplacians on the Weyl Chamber}\label{sec3}
In this section we associate a commuting system of discrete pseudo
Laplacians on $\mathcal{P}^+$  to our orthonormal polynomials
$P_\lambda (\xi)$.

\subsection{Fourier Transform}\label{sec3.1}
Let $\mathcal{H}$ be the Hilbert space $ l^2(\mathcal{P}^+)$ of
square-summable functions over the dominant cone $\mathcal{P}^+$
equipped with the standard inner product
\begin{subequations}
\begin{equation}
(f,g)_{\mathcal{H}}=\sum_{\lambda\in\mathcal{P}^+} f_\lambda
\overline{g_\lambda}\qquad (f,g\in l^2 (\mathcal{P}^+)),
\end{equation}
and let $\hat{\mathcal{H}}$ be the Hilbert space $L^2
(\mathbf{A},\text{d}\xi)$ of square-integrable functions over the
Weyl alcove equipped with the normalized inner product
\begin{equation}
(\hat{f},\hat{g})_{\hat{\mathcal{H}}}=\frac{1}{|W|\,\text{Vol}(\mathbf{A})}\int_{\mathbf{A}}
\hat{f}(\xi) \overline{\hat{g}(\xi)} \text{d}\xi \qquad
(\hat{f},\hat{g}\in L^2 (\mathbf{A},\text{d}\xi)).
\end{equation}
\end{subequations}
By construction, the functions
\begin{equation}\label{wave-f}
\Psi_\lambda (\xi) =  \hat{\Delta}^{1/2} (\xi) \delta (\xi)
P_\lambda (\xi) , \qquad \lambda\in\mathcal{P}^+
\end{equation}
form an orthonormal basis of $\hat{\mathcal{H}}$. As a
result, the mapping
$\mathcal{F}:\mathcal{H}\mapsto\hat{\mathcal{H}}$ given by
$\phi_\lambda
\stackrel{\mathcal{F}}{\longrightarrow}\hat{\phi}(\xi)$ with
\begin{subequations}
\begin{eqnarray}\label{F1}
\hat{\phi} (\xi) &=& (\phi ,\Psi (\xi))_{\mathcal{H}} \\
&=& \sum_{\lambda\in\mathcal{P}^+} \phi_\lambda
\overline{\Psi_\lambda (\xi )}   \nonumber
\end{eqnarray}
constitutes a unitary Hilbert space isomorphism between $\mathcal{H}$ and
$\hat{\mathcal{H}}$. The inverse mapping
$\mathcal{F}^{-1}:\hat{\mathcal{H}}\mapsto\mathcal{H}$ takes the
form $\hat{\phi}(\xi)\stackrel{\mathcal{F}^{-1}}{\longrightarrow}
\phi_\lambda$ with
\begin{eqnarray}\label{F2}
\phi_\lambda &=& (\hat{\phi}
,\overline{\Psi}_\lambda)_{\hat{\mathcal{H}}} \\
&=& \frac{1}{|W|\,\text{Vol}(\mathbf{A})}\int_{\mathbf{A}}
\hat{\phi}(\xi) \Psi_\lambda(\xi) \text{d}\xi. \nonumber
\end{eqnarray}
\end{subequations}
We will refer to $\mathcal{F}$ as the {\em Fourier transform}
associated to the polynomials $P_\lambda (\xi)$. In the simplest
case with unit $c$-functions, the wave functions amount to plane
waves (cf. Section \ref{weyl:sec})
\begin{equation}\label{pwaves}
\Psi_\lambda^{(0)} (\xi) = \sum_{w\in W} (-1)^w\, e^{i\langle
\rho+\lambda , \xi_w\rangle } .
\end{equation}
The corresponding Fourier transform reduces to the conventional
Fourier transform
$\mathcal{F}^{(0)}:\mathcal{H}\mapsto\hat{\mathcal{H}}$ of the
form $\phi_\lambda
\stackrel{\mathcal{F}^{(0)}}{\longrightarrow}\hat{\phi} (\xi)$
with
\begin{subequations}
\begin{equation}\label{F01}
\hat{\phi} (\xi) =  \sum_{w\in W} (-1)^w\,
\sum_{\lambda\in\mathcal{P^+}}
 \phi_\lambda e^{-i\langle \rho+\lambda , \xi_w\rangle } .
\end{equation}
The inverse transform
$(\mathcal{F}^{(0)})^{-1}:\hat{\mathcal{H}}\mapsto\mathcal{H}$ is
then given by $\hat{\phi} (\xi)
\stackrel{(\mathcal{F}^{(0)})^{-1}}{\longrightarrow} \phi_\lambda$
with
\begin{equation}\label{F02}
\phi_\lambda = \frac{1}{|W|\,\text{Vol} (\mathbf{A})} \sum_{w\in
W} (-1)^w\, \int_{\mathbf{A}} \hat{\phi} (\xi) e^{i\langle
\rho+\lambda , \xi_w\rangle } \text{d}\xi .
\end{equation}
\end{subequations}

\subsection{Pseudo Laplacians}\label{sec3.2}
To the basis of fundamental weights $\omega_1,\ldots ,\omega_N$
generating $\mathcal{P}^+$, we associate bounded
multiplication operators $\hat{E}_1,\ldots ,\hat{E}_N$ in
$\hat{\mathcal{H}}$ of the form
\begin{equation}
\hat{E}_r(\xi) = \sum_{\nu\in  W(\omega_r)} \exp
( i \langle \nu , \xi \rangle ), \qquad r=1,\ldots ,N,
\end{equation}
where the sum is over all weights in the Weyl orbit of $\omega_r$.
The pullbacks of $\hat{E}_1,\ldots ,\hat{E}_N$ with respect to the
Fourier transform $\mathcal{F}$ define an integrable system of bounded
commuting operators in $\mathcal{H}$
\begin{equation}\label{pL}
L_r = \mathcal{F}^{-1}\circ \hat{E}_r \circ \mathcal{F}, \qquad
r=1,\ldots ,N.
\end{equation}
We will refer to the commutative algebra $\mathbb{R}[L_1,\ldots
,L_N]$ generated by these operators as the (algebra of) {\em
discrete pseudo Laplacians} associated to the polynomials
$P_\lambda (\xi)$.
It is immediate from its construction as the pullback of a
multiplication operator in $\hat{\mathcal{H}}$ (cf. Eq. \eqref{pL}) that
the pseudo Laplacian $L_r$ has a purely absolutely
continuous spectrum in $\mathcal{H}$ given by the compact set
$\sigma (L_r)=\{ \hat{E}_r(\xi) \mid
\xi\in\overline{\mathbf{A}}\}\subset\mathbb{C}$.
By acting with both sides of the operator equality
$L_r \mathcal{F}^{-1}= \mathcal{F}^{-1} \hat{E}_r$
on (the complex conjugate of)
an arbitrary element $\hat{\phi}\in\hat{\mathcal{H}}$, we get
\begin{subequations}
\begin{equation}\label{ev-eqr}
L_r(\Psi_\lambda ,\hat{\phi})_{\hat{\mathcal{H}}}=
(\hat{E}_r\Psi_\lambda,\hat{\phi})_{\hat{\mathcal{H}}},
\qquad\forall\hat{\phi}\in\hat{\mathcal{H}}.
\end{equation}
In other words,
the functions $\Psi_\lambda (\xi)$ form a complete (as
$\mathcal{F}:\mathcal{H}\rightarrow\hat{\mathcal{H}}$
is a Hilbert space isomorphism) set of generalized joint
eigenfunctions of our pseudo Laplacians, i.e. formally
\begin{equation}\label{ev-eq}
L_r \Psi_\lambda (\xi ) = \hat{E}_r (\xi) \Psi_\lambda (\xi) .
\end{equation}
\end{subequations}
Here $\xi\in\overline{\mathbf{A}}$ plays the role of the spectral parameter
and the weight $\lambda\in\mathcal{P}^+$ is interpreted as the
discrete geometric variable (i.e. the position variable).

\begin{note}
{\em i.} Below we will sometimes write formal
equalities of the form in Eq. \eqref{ev-eq} that admit a
rigorous interpretation of the form in Eq. \eqref{ev-eqr} upon taking the
inner product (smearing)
with an arbitrary (stationary) wave packet $\hat{\phi}\in\hat{\mathcal{H}}$.
\end{note}

\begin{note}
{\em ii.}
In general the Laplacian $L_r$ is not self-adjoint. Indeed, the adjoint
$L_r^*$ is given by  $L_s$ with $\omega_s=-w_0(\omega_r)$, where
$w_0$ denotes the longest element of the Weyl group $W$ (i.e., the
unique Weyl group element $w_0$ such that $w_0(\mathbf{A})=-\mathbf{A}$).
Thus $L_r$ is self-adjoint if and only if $w_0(\omega_r)=-\omega_r$.
\end{note}

\subsection{Localization}\label{sec3.3}
Let $\phi:\mathcal{P}^+\rightarrow\mathbb{C}$ be square-summable a
lattice function. The action of $L_r$ on $\phi$ is of the form
\begin{equation}
L_r \phi_\lambda = \sum_{\mu\in\mathcal{P}^+} a_{\lambda\mu ;r}
\phi_\mu,
\end{equation}
for certain coefficients  $a_{\lambda\mu ;r}\in \mathbb{C}$. We
will now show that in fact only a finite number of these
coefficients is nonzero.

\begin{proposition}[Localization]\label{localization:prp} The action of the
pseudo
Laplacian $L_r$ on $\phi\in\mathcal{H}$ is of the form
\begin{subequations}
\begin{equation}
L_r \phi_\lambda = \sum_{\mu\in \mathcal{P}^+_{\lambda ;r}}
a_{\lambda\mu ;r} \phi_\mu ,
\qquad a_{\lambda\mu ;r}\in \mathbb{C},
\end{equation}
where
\begin{equation}\label{locset}
\mathcal{P}^+_{\lambda ;r} =\{ \mu\in\mathcal{P}^+ \mid \mu\preceq
\lambda +\omega_r \; \text{and}\; \mu-w_0(\omega_r)\succeq \lambda
\} .
\end{equation}
\end{subequations}
\end{proposition}
\begin{proof}
From the triangularity of the monomial expansion of $P_\lambda
(\xi)$ it is immediate that
\begin{equation*}
m_{\omega_r}(\xi) P_\lambda (\xi) = \sum_{\begin{subarray}{c}
\mu\in\mathcal{P}^+\\ \mu\preceq \lambda+\omega_r\end{subarray}}
a_{\lambda\mu
;r } P_\mu (\xi ) ,\qquad a_{\lambda\mu ;r}\in \mathbb{C}.
\end{equation*}
The orthonormality furthermore implies that
\begin{equation*}
a_{\lambda\mu ;r }=(m_{\omega_r} P_\lambda ,
P_\mu)_{\hat{\Delta}}= ( P_\lambda ,
m_{\omega_s}P_\mu)_{\hat{\Delta}}=\overline{a_{\mu\lambda ;s
}} ,
\end{equation*}
with $\omega_s=-w_0(\omega_r)$ (cf. Note {\em ii}. above).
Hence
\begin{equation*}
a_{\lambda\mu ;r }\neq 0\Rightarrow \mu\in\mathcal{P}^+_{\lambda
;r} .
\end{equation*}
Since $\hat{E}_r(\xi)=m_{\omega_r}(\xi)$ and
$ \Psi_\lambda (\xi )=\hat{\Delta}^{1/2}(\xi)\delta(\xi)P_\lambda(\xi)$,
we conclude that
\begin{equation*}
\hat{E}_r (\xi) \Psi_\lambda (\xi) = \sum_{\mu\in
\mathcal{P}^+_{\lambda ;r}  }
a_{\lambda\mu ;r}
 \Psi_\mu (\xi ) .
\end{equation*}
Taking the innerproduct with an arbitrary wave packet
$\hat{\phi}\in\hat{\mathcal{H}}$ and
comparison with the eigenvalue equation in Eq.
\eqref{ev-eqr} entails that
\begin{equation*}
L_r (\Psi_\lambda ,\hat{\phi})_{\hat{\mathcal{H}}}=
\sum_{\mu\in \mathcal{P}^+_{\lambda ;r}}
a_{\lambda\mu ;r} (\Psi_\mu ,\hat{\phi})_{\hat{\mathcal{H}}} ,
\qquad \forall\hat{\phi}\in\hat{\mathcal{H}},
\end{equation*}
i.e. formally (cf. Note {\em i.} above)
\begin{equation*}
L_r \Psi_\lambda (\xi )= \sum_{\mu\in \mathcal{P}^+_{\lambda ;r}}
a_{\lambda\mu ;r} \Psi_\mu (\xi ) .
\end{equation*}
The proposition then follows by the completeness of the generalized
eigenfunctions $\Psi_\lambda (\xi )$, $\xi \in \overline{\mathbf{A}}$ in the
Hilbert space $\mathcal{H}$ (i.e. by the fact that the Fourier transform
$\mathcal{F}$
\eqref{F1}, \eqref{F2} constitutes a
unitary Hilbert space isomorphism between
$\mathcal{H}$ and $\hat{\mathcal{H}}$).
\end{proof}

A priori the cardinality of the set $\mathcal{P}^+_{\lambda
;r}$ may be unbounded as a
function  of $\lambda\in\mathcal{P}^+$. Hence, in general our
pseudo Laplacians need {\em not} be difference operators. If the
ordering of the dominant weights $\succeq$ coincides with the
dominance order $\geqslant$, however, then it follows from the
definition in Eq. \eqref{po} that the size of the set
$\mathcal{P}^+_{\lambda ;r}$ is
bounded by the number of
weights in the interval $\{ \nu \in\mathcal{P} \mid w_0(\omega_r)
\leqslant \nu \leqslant \omega_r \}$. Consequently, in this
situation our pseudo Laplacian $L_r$ is actually a difference
operator in $\mathcal{H}$. (We will refer in such case to $L_r$ as
a {\em discrete Laplacian} as opposed to merely a pseudo
Laplacian.)

\begin{proposition}[Discrete Laplacians]\label{dl:prp}
When our ordering $\succeq$ coincides with the dominance ordering
$\geqslant$ \eqref{po}, then the pseudo Laplacians in
$\mathbb{R}[L_1,\ldots ,L_N]$ are discrete difference operators in
$\mathcal{H}$.
\end{proposition}

In the case of unit $c$-functions (cf. Section \ref{weyl:sec}),
our discrete Laplacians $L_1,\ldots ,L_N$ amount to conventional
free Laplacians $L_1^{(0)},\ldots ,L_N^{(0)}$ over the dominant
cone $\mathcal{P}^+$.

\begin{proposition}[Free Laplacians]\label{fl:prp}
If $\hat{c}_{|\alpha |}(z)=1$, $\forall \alpha\in
\boldsymbol{R}_1^+$, then our discrete Laplacians $L_r$ reduce to
the free Laplacians
\begin{equation*}
L_r^{(0)} \phi_\lambda = \sum_{\nu\in W(\omega_r)}
\phi_{\lambda +\nu} , \qquad r=1,\ldots ,N,
\end{equation*}
with the boundary condition that for
$\mu\in\mathcal{P}\setminus\mathcal{P}^+$
\begin{equation*}
\phi_\mu  =
\begin{cases}
(-1)^{w_{\rho+\mu}} \phi_{w_{\rho+\mu} (\rho+\mu)-\rho}
&\text{if}\;\; |W_{\rho+\mu}| =1,  \\
0& \text{if}\;\; |W_{\rho+\mu}| >1
\end{cases}
\end{equation*}
(where $w_{\rho+\mu}$ denotes the Weyl permutation taking the
regular weight $\rho+\mu$ to the dominant cone).
\end{proposition}
\begin{proof}
As pointed out in Section \ref{weyl:sec}, the case of unit
$c$-functions corresponds to orthonormal polynomials $P_\lambda
(\xi)$ given by the Weyl characters $ \chi_\lambda (\xi)$. It is
immediate from the explicit expression for $\chi_\lambda$ in Eq.
\eqref{weylcars} that the Weyl characters satisfy the well-known
recurrence
relations
\begin{equation*}
m_{\omega_r}(\xi) \chi_\lambda (\xi) = \sum_{\nu\in W(\omega_r)}
\chi_{\lambda+\nu} (\xi ) .
\end{equation*}
Starting from these recurrence relations, the proposition readily
follows by repeating the arguments in the proof of Proposition
\ref{localization:prp}. The boundary condition stems from the
property  \eqref{weylcars2} of the Weyl characters.
\end{proof}

By Proposition \ref{localization:prp}, the eigenvalue equations in
Eq.  \eqref{ev-eq} take the form
\begin{subequations}
\begin{equation}
\sum_{\mu\in \mathcal{P}^+_{\lambda
;r}  } a_{\lambda\mu ;r}
 \Psi_\mu (\xi ) = \hat{E}_r (\xi) \Psi_\lambda (\xi), \qquad 1,\ldots
 ,N,
\end{equation}
or equivalently
\begin{equation}
\sum_{\mu\in \mathcal{P}^+_{\lambda
;r}  } a_{\lambda\mu ;r}
 P_\mu (\xi ) = \hat{E}_r (\xi) P_\lambda (\xi), \qquad 1,\ldots
 ,N
\end{equation}
\end{subequations}
(upon dividing out the trivial overall normalization factor
$\delta (\xi)/
\sqrt{\hat{\mathcal{C}}(\xi)\hat{\mathcal{C}}(-\xi)}$ on both
sides). The latter equations admit an alternative interpretation
as a system of recurrence relations (or Pieri formulas) for the
polynomials $P_\lambda (\xi)$.

\begin{note}
The above construction of the discrete (pseudo) Laplacians has its origin in
the works of Macdonald \cite{mac:spherical,mac:orthogonal,mac:affine}.
Specifically,
for $c_{|\alpha |}(z)=(1-t_{|\alpha|}z)$ with $-1<t_{|\alpha |}<1$ the
polynomials $P_\lambda (\xi)$ \eqref{op1}, \eqref{op2}
amount to (the parameter deformations of) Macdonald's
zonal spherical functions on $p$-adic Lie groups
\cite{mac:spherical,mac:orthogonal}. The algebra
of discrete Laplacians $\mathbb{R}[L_1,\ldots ,L_N]$ corresponds in this case
to the $K$-spherical Hecke algebra of the $p$-adic Lie group.
When  $c_{|\alpha |}(z)$ is given by a $q$-shifted factorial
(cf. Eq. \eqref{m-c} below), then the
polynomials $P_\lambda (\xi)$ specialize to the Macdonald polynomials
\cite{mac:symmetric,mac:orthogonal,mac:affine}.
The discrete Laplacians appear in this context in Cherednik's
double affine Hecke algebra as ``coordinate multiplication operators''
dual to Macdonald's difference operators \cite{che:macdonalds,mac:affine}.
\end{note}

\section{Time-Dependent Scattering Theory}\label{sec4}
In this section we determine the wave operators and scattering
operator associated to our discrete pseudo Laplacians. For
background literature on scattering theory the reader is referred
to e.g. Refs. \cite{ree-sim:methods,pea:quantum,thi:course}.

\subsection{Plane Wave Asymptotics}\label{sec4.1}
The dominant Weyl chamber is given by the open convex cone
\begin{equation}\label{dwc}
\mathbf{C}^+=\{ \mathbf{x}\in \mathbf{E}\mid \langle \mathbf{x},
\alpha \rangle
> 0,\; \forall \alpha\in \boldsymbol{R}^+ \} .
\end{equation}
We will now describe the asymptotics of the wave function
$\Psi_\lambda (\xi)$
\eqref{wave-f} diagonalizing the pseudo Laplacians $L_1,\ldots
,L_N$ \eqref{pL} for $\lambda$ deep in the Weyl chamber, i.e.,
for $\lambda$ growing to infinity in such a way that
$\langle \lambda ,\alpha^\vee\rangle \to +\infty$ for all
positive roots
$\alpha\in\boldsymbol{R}^+$.

To this end we define
for $\lambda\in\mathcal{P}^+$
\begin{equation}
m(\lambda) \equiv \min_{\alpha\in \boldsymbol{R}^+} \langle
\lambda ,\alpha^\vee\rangle .
\end{equation}
In previous work, it was shown that the strong $L^2$-asymptotics
of the polynomials $P_\lambda (\xi)$ for
$m(\lambda)\to\infty$ is
given by \cite{rui:factorized,die:asymptotic,die:asymptotics}
\begin{equation}
P_\lambda^\infty (\xi) = \delta^{-1}(\xi) \sum_{w\in W} (-1)^w
\hat{\mathcal{C}}(\xi_w) e^{i\langle \rho +\lambda , \xi_w
\rangle}.
\end{equation}
More precisely, one has that
\begin{equation}\label{pol-as}
\| P_\lambda -P_\lambda^\infty \|_{\hat{\Delta}} =
O(e^{-\epsilon\, m(\lambda) })\quad \text{as}\;\;
 m(\lambda)\longrightarrow \infty ,
\end{equation}
where $\| \cdot \|_{\hat{\Delta}} \equiv (\cdot ,\cdot
)_{\hat{\Delta}}^{1/2}$ and $\epsilon>0$ denotes a decay rate that
depends on the radius $\varrho>1$ of the analyticity disc
$\mathbb{D}_\varrho$ of the $c$-functions $\hat{c}_{|\alpha |}(z)$
(see the technical assumptions in Section \ref{sec2}).

\begin{subequations}
The idea of the proof in
\cite{rui:factorized,die:asymptotic,die:asymptotics} of this
exponential convergence goes along the following lines. Firstly, a
direct (constant term) computation reveals that
\begin{equation}\label{prf1}
\langle P_\lambda^{\infty} ,m_\mu\rangle_{\hat{\Delta}} =
\begin{cases}
0& \text{if}\; \mu\prec\lambda , \\
1& \text{if}\: \mu = \lambda .
\end{cases}
\end{equation}
Next, we denote by $P_\lambda^{(m(\lambda ))}(\xi) $ the
polynomial approximation of the asymptotic function
$P_\lambda^\infty (\xi)$ obtained by replacing the overall
$c$-function $\hat{\mathcal{C}}(\xi)$ by its Taylor polynomial of
degree $m(\lambda)$.  Then a combinatorial analysis shows that
this polynomial approximation expands triangularly on the basis of
monomial symmetric functions
\begin{equation}\label{prf2}
P_\lambda^{(m(\lambda ))}(\xi)=m_\lambda
(\xi)+\sum_{\mu\in\mathcal{P}^+,\mu\prec\lambda} b_{\lambda\mu}
m_\mu (\xi)
\end{equation}
(for certain coefficients $b_{\lambda\mu}\in\mathbb{C}$).
Moreover, the analyticity requirements on the $c$-functions
$\hat{c}_{|\alpha |}(z)$ guarantee that
\begin{equation}\label{prf3}
P_\lambda^\infty (\xi) = P_\lambda^{m(\lambda )}(\xi) +
O(e^{-\epsilon m(\lambda )})
\end{equation}
(this is because the technical conditions ensure
that the Taylor coefficients of the $c$-function
$\hat{c}_{|\alpha |}(z)$ decay exponentially fast).
\end{subequations}
From Eqs. \eqref{prf1}-\eqref{prf3} one concludes that---up to an
$O(e^{-\epsilon m(\lambda )})$ error term---the asymptotic
function $P_\lambda^\infty(\xi)$ amounts to a monic polynomial
obtained by performing the Gram-Schmidt process on the monomial
symmetric basis with respect to the inner product $(\cdot ,\cdot
)_{\hat{\Delta}}$. In other words, the asymptotic functions
coincide up to exponentially decaying error terms with
the monic versions of the polynomials
$P_\lambda (\xi)$ defined in Eqs. \eqref{op1}, \eqref{op2}. The
convergence in Eq. \eqref{pol-as} now follows from the fact that the
orthonormalized polynomials $P_\lambda(\xi)$ are asymptotically
monic: $a_{\lambda\lambda}=1+O(e^{-\epsilon
m(\lambda)})$. (This estimate for the leading coefficient
in the monomial expansion of $P_\lambda (\xi)$
follows starting from the equality
$a_{\lambda\lambda} =\langle P_\lambda
,P_\lambda^\infty\rangle_{\hat{\Delta}}$, upon substituting
\eqref{prf3} and expanding the polynomial part $P_\lambda^{m(\lambda)}(\xi)$
in terms of the
orthonormalized polynomials $P_\mu(\xi)$, $\mu\preceq\lambda$, taking into
account the orthogonality \eqref{op2}.)

The asymptotic estimate in Eq. \eqref{pol-as} for the polynomials
$P_\lambda (\xi)$
immediately gives rise to the following plane wave
asymptotics for the wave functions $\Psi_\lambda (\xi )$ \eqref{wave-f}:
\begin{subequations}
\begin{eqnarray}\label{aswave}
\Psi_\lambda^\infty (\xi)  &= & \hat{\Delta}^{1/2}(\xi) \delta
(\xi) P_\lambda^\infty
(\xi)  \\
&=& \sum_{w\in W} (-1)^w \hat{S}_w^{1/2}(\xi) e^{i\langle \rho
+\lambda , \xi_w \rangle} ,
\end{eqnarray}
where
\begin{eqnarray}
\hat{S}_w (\xi) &=& \frac{\hat{\mathcal{C}}(\xi_w)}{\hat{\mathcal{C}}(-\xi_w)} \\
&=& \prod_{\alpha\in \boldsymbol{R}^+_1\cap
w^{-1}(\boldsymbol{R}^+_1)} \hat{s}_{|\alpha|}(\langle \alpha ,
\xi\rangle) \prod_{\alpha\in \boldsymbol{R}^+_1\cap
w^{-1}(-\boldsymbol{R}^+_1)} \overline{\hat{s}_{|\alpha|}(\langle
\alpha , \xi\rangle)} , \label{Sw}
\end{eqnarray}
with
\begin{equation}\label{smat}
\hat{s}_{|\alpha|}(\langle \alpha , \xi\rangle)=
\frac{\hat{c}_{|\alpha|}(e^{-i\langle \alpha ,
\xi\rangle})}{\hat{c}_{|\alpha|}(e^{i\langle \alpha ,
\xi\rangle})}
\end{equation}
\end{subequations}
(so $\overline{\hat{s}_{|\alpha|}(\langle \alpha ,
\xi\rangle)}=\hat{s}_{|\alpha|}(-\langle \alpha ,
\xi\rangle)=\hat{s}_{|\alpha|}^{-1}(\langle \alpha , \xi\rangle)$
and $|\hat{s}_{|\alpha|}(\langle \alpha , \xi\rangle)|=1$ ).

\begin{theorem}[Plane Wave Asymptotics]\label{plane:thm}
The wave function $\Psi_\lambda$ tends to the plane waves
$\Psi_\lambda^\infty$ for $\lambda$ deep in the Weyl chamber:
\begin{equation*}
\| \Psi_\lambda -\Psi_\lambda^\infty \|_{\hat{\mathcal{H}}} =
O(e^{-\epsilon\, m(\lambda) })\quad \text{as}\;\;
m(\lambda)\longrightarrow \infty ,
\end{equation*}
where $\| \cdot \|_{\hat{\mathcal{H}}}\equiv (\cdot
,\cdot)_{\hat{\mathcal{H}}}^{1/2}$.
\end{theorem}

We see from Theorem \ref{plane:thm} that the asymptotics of the
wave functions $\Psi_\lambda (\xi)$ for $\lambda$ deep in the Weyl
chamber is given by an anti-symmetric combination of plane waves
$e^{i\langle \lambda ,\xi \rangle}$ with phase-factors that
factorize over the root system in terms of one-dimensional
$c$-functions.

\subsection{Scattering and Wave Operators}\label{sec4.2}
For any {\em real} multiplication operator $\hat{E}(\xi) \subset
\mathbb{R}[\hat{E}_1 (\xi),\ldots ,\hat{E}_N(\xi)]$, let $L=
\mathcal{F}^{-1}\circ \hat{E} \circ\mathcal{F}$ and let $L^{(0)}=
(\mathcal{F}^{(0)})^{-1}\circ \hat{E} \circ\mathcal{F}^{(0)}$.  In
other words, the operators $L\subset \mathbb{R}[L_1,\ldots ,L_N]$
and $L^{(0)}\subset \mathbb{R}[L_1^{(0)},\ldots ,L_N^{(0)}]$ are
{\em self-adjoint} (pseudo) Laplacians in $\mathcal{H}$ such that
(formally)
\begin{equation}
L \Psi_\lambda (\xi ) = \hat{E} (\xi) \Psi_\lambda (\xi) \quad
\text{and}\quad L^{(0)} \Psi_\lambda^{(0)} (\xi ) = \hat{E} (\xi)
\Psi_\lambda^{(0)} (\xi) .
\end{equation}
(So the spectrum of $L$ and $L^{(0)}$ in $\mathcal{H}$ is
absolutely continuous and given by the compact interval $\sigma
(L)=\sigma (L^{(0)})=\{ \hat{E}(\xi) \mid
\xi\in\overline{\mathbf{A}}\}$.) We will now describe the
scattering of the interacting dynamics generated by the discrete
pseudo Laplacian $L$ with respect to the free dynamics generated
by the discrete Laplacian $L^{(0)}$. Let us to this end define the
regular sector of the Weyl alcove as
\begin{equation}
\mathbf{A}_{\text{reg}}=\{ \xi\in \mathbf{A}\mid \langle \nabla
\hat{E}, \alpha \rangle \neq 0,\; \forall \alpha\in
\boldsymbol{R}^+ \} .
\end{equation}
Due to the analyticity of $\hat{E}(\xi)$, the regular sector
$\mathbf{A}_{\text{reg}}$ is an open dense subset of the Weyl
alcove $\mathbf{A}$. For every $\xi \in\mathbf{A}_{\text{reg}}$,
there exists now a unique Weyl group element $\hat{w}_\xi\in W$
such that $\hat{w}_\xi (\nabla \hat{E})$ lies in the dominant Weyl
chamber $ \mathbf{C}^+$ \eqref{dwc}. Clearly, the Weyl-group valued
function $\xi\rightarrow \hat{w}_\xi$ is constant on the connected
components of $\mathbf{A}_{\text{reg}}$ (by continuity). We are
now in the position to define the unitary multiplication operator
$\hat{\mathcal{S}}_L:\hat{\mathcal{H}}\mapsto\hat{\mathcal{H}}$
(the so-called {\em scattering matrix}) via its restriction to the
dense subspace of (say) smooth complex test functions with compact
support in $\mathbf{A}_{\text{reg}}$:
\begin{equation}\label{Sm}
(\hat{\mathcal{S}}_L\hat{\phi} ) (\xi)=
\hat{S}_{\hat{w}_\xi}(\xi)\hat{\phi} (\xi) \qquad ( \hat{\phi}\in
C_0^\infty (\mathbf{A}_{\text{reg}}) ),
\end{equation}
where $\hat{S}_{w}(\xi)$ is given by Eq. \eqref{Sw}.

The main result of this paper is the following explicit formula
for the wave operators and the scattering operator in terms of the
scattering matrix $\hat{\mathcal{S}}_L$ \eqref{Sm} and the Fourier
transforms $\mathcal{F}$ \eqref{F1}, \eqref{F2} and
$\mathcal{F}^{(0)}$ \eqref{F01}, \eqref{F02}, thus relating the
long-time asymptotics of {\em interacting dynamics} $e^{itL}$ to
that of the {\em free dynamics} $e^{itL^{(0)}}$. The proof, which
is relegated to Section \ref{sec5} below, consists of a stationary
phase analysis based on the asymptotic formula for the wave
functions in Theorem \ref{plane:thm}.

\begin{theorem}[Wave Operators]\label{wave:thm}
The operator limits
\begin{equation*}
\Omega_\pm = s-\lim_{t\to\pm\infty} e^{itL}e^{-itL^{(0)}}
\end{equation*}
converge in the strong $\|\cdot \|_{\mathcal{H}}$-norm topology (where
$\|\cdot \|_{\mathcal{H}}=(\cdot,\cdot)_{\mathcal{H}}^{1/2}$),
and the corresponding wave operators
$\Omega_\pm:\mathcal{H}\mapsto\mathcal{H}$ are given by the
unitary operators
\begin{eqnarray*}
\Omega_+ &=& \mathcal{F}^{-1}\circ\hat{\mathcal{S}}_L^{-1/2}\circ
\mathcal{F}^{(0)} , \\
\Omega_- &=& \mathcal{F}^{-1}\circ\hat{\mathcal{S}}_L^{1/2}\circ
\mathcal{F}^{(0)} .
\end{eqnarray*}
\end{theorem}

\begin{corollary}[Scattering Operator]\label{scattering:cor}
The scattering operator
$\mathcal{S}_L:\mathcal{H}\mapsto\mathcal{H}$ for the self-adjoint
discrete
pseudo Laplacian $L\in \mathbb{R}[L_1,\ldots ,L_N]$ is given by
the unitary operator
\begin{equation*}
\mathcal{S}_L
\equiv\Omega_+^{-1}\Omega_-=(\mathcal{F}^{(0)})^{-1}\circ\hat{\mathcal{S}}_L\circ
\mathcal{F}^{(0)} .
\end{equation*}
\end{corollary}

We see from Corollary \ref{scattering:cor} and Eqs. \eqref{Sw},
\eqref{Sm} that the scattering matrix for the self-adjoint
discrete pseudo
Laplacian $L$ factorizes over the root system $\boldsymbol{R}$.
For the type $A$ root systems,  Theorem \ref{wave:thm} and
Corollary \ref{scattering:cor} reproduce the results of
Ruijsenaars in Ref. \cite{rui:factorized}.

\section{Stationary Phase Analysis}\label{sec5}
In this section the fundamental formulas for the wave operators
stated in Theorem \ref{wave:thm} are proven. To this end we employ
a stationary phase method that generalizes Ruijsenaars' approach
in Ref. \cite{rui:factorized} from the type $A$ root systems to
the case of arbitrary crystallographic root systems. Throughout
this section the notational conventions of Sections \ref{sec3} and
\ref{sec4} are adopted.

\subsection{Asymptotics of Wave Packets}\label{sec5.1}
Let us introduce the {\em free wave packet} $\phi^{(0)}(t) $ and
the {\em interacting wave packets} $\phi_{\pm }(t)$ of the form
\begin{subequations}
\begin{eqnarray}
\phi^{(0)}(t) &=& (\mathcal{F}^{(0)})^{-1}\, e^{-i
t\hat{E}}\,\hat{\phi}   , \\
\phi_{\pm }(t) &=& \mathcal{F}^{-1}\, e^{-i
t\hat{E}}\,\hat{\mathcal{S}}_L^{\mp 1/2}\, \hat{\phi},
\end{eqnarray}
\end{subequations}
or more explicitly
\begin{subequations}
\begin{eqnarray}
 \phi^{(0)}_\lambda (t) &=&
\frac{1}{|W|\,\text{Vol} (\mathbf{A})} \sum_{w\in W} (-1)^w\,
\int_{\mathbf{A}} e^{i\langle \rho+\lambda ,
\xi_w\rangle-it\hat{E}(\xi) } \hat{\phi} (\xi) \text{d}\xi , \\
\phi_{\pm,\lambda} (t)&=& \frac{1}{|W|\text{Vol}(\mathbf{A})}
\int_{\mathbf{A}} \Psi_\lambda (\xi) e^{-it\hat{E}(\xi)}
\hat{\mathcal{S}}_L^{\mp 1/2} (\xi) \hat{\phi}(\xi) \text{d}\xi ,
\end{eqnarray}
\end{subequations}
with $\hat{\phi}\in C_0^\infty (\mathbf{A}_{\text{reg}})$. The
following lemma states that the long-time asymptotics of the
interacting wave packets $\phi_{+ }(t)$ and $\phi_{- }(t)$ for
$t\to +\infty$ and $t\to -\infty$, respectively, coincides with
the corresponding asymptotics of the free wave packet
$\phi^{(0)}(t)$.

\begin{proposition}[Asymptotic Freedom]\label{asf:prp}
For $t\to\pm \infty$, the difference between the interacting wave
packet $\phi_{\pm }(t)$ and the free wave packet $\phi^{(0)}(t)$
tends to zero:
\begin{equation*}
\forall \kappa >0:\qquad \| \phi_{\pm }(t)- \phi^{(0)}(t)
\|_{\mathcal{H}} =O(1/|t|^\kappa)\quad \text{as}\;t\to \pm\infty .
\end{equation*}
\end{proposition}

Before proving Proposition \ref{asf:prp} (cf. below), let us first
infer that Theorem \ref{wave:thm} arises as an immediate
consequence. Indeed, since the space of test functions
$C_0^{\infty}(\mathbf{A}_{\text{reg}})$ is dense in
$\hat{\mathcal{H}}$ and the operators in question are unitary, to
validate Theorem \ref{wave:thm} it is sufficient to demonstrate
that for $\phi = (\mathcal{F}^{(0)})^{-1}\hat{\phi}$ with
$\hat{\phi}\in C_0^{\infty}(\mathbf{A}_{\text{reg}})$
\begin{equation*}
\lim_{t\to\pm\infty} \| e^{itL}e^{-itL^{(0)}}\phi-\Omega_{\pm}
\phi \|_{\mathcal{H}} =0,
\end{equation*}
where $\Omega_{\pm}\equiv\mathcal{F}^{-1}\circ
\hat{\mathcal{S}}_L^{\mp 1/2}\circ \mathcal{F}^{(0)}$. From the
unitarity of $e^{itL}$ and the intertwining relations
\begin{equation*}
e^{-itL^{(0)}} \circ (\mathcal{F}^{(0)})^{-1} =
(\mathcal{F}^{(0)})^{-1}\circ e^{-it\hat{E}}\quad\text{and} \quad
e^{-itL} \circ\mathcal{F}^{-1} = \mathcal{F}^{-1}\circ
e^{-it\hat{E}},
\end{equation*}
it is clear that
\begin{eqnarray*}
\| e^{itL}e^{-itL^{(0)}}\phi-\Omega_{\pm} \phi \|_{\mathcal{H}}
&=& \| e^{-itL^{(0)}}(\mathcal{F}^{(0)})^{-1}\hat{\phi}-
e^{-itL}\Omega_{\pm} (\mathcal{F}^{(0)})^{-1}\hat{\phi} \|_{\mathcal{H}}\\
 &=& \| \phi^{(0)}(t)- \phi_{\pm }(t) \|_{\mathcal{H}},
\end{eqnarray*}
whence the theorem follows from Proposition \ref{asf:prp}.

\subsection{The Classical Wave Packet}\label{sec5.2}
To prove Proposition \ref{asf:prp}, we may assume without
restricting generality that the test function $\hat{\phi}$ has in
fact compact support inside a connected component of
$\mathbf{A}_{\text{reg}}$. In this situation there thus exists a
unique Weyl group element $\hat{w}\in W$ such that
$\hat{w}(\nabla\hat{E})$ lies inside the dominant Weyl chamber
$\mathbf{C}^+$ \eqref{dwc} for all $\xi$ in the support of $\hat{\phi}$. Let
$\mathbf{V}_{\text{clas}}\subset\mathbf{E}$ be an open bounded
neighborhood of the compact range of classical wave-packet
velocities $\text{Ran}_{\hat{\phi}} (\nabla \hat{E}) \equiv\{
\nabla\hat{E}(\xi) \mid \xi \in \text{Supp}(\hat{\phi})\}$ staying
away from the walls of the Weyl chamber
$\hat{w}^{-1}(\mathbf{C}^+)$ in the sense that there exists a
lower-bound $\varepsilon>0$ such that $\langle \zeta_{\hat{w}}
,\alpha^\vee \rangle
>\varepsilon$ for all $\zeta\in\mathbf{V}_{\text{clas}}$ and all $\alpha\in \boldsymbol{R}^+$.
We will now introduce a {\em classical wave packet} that is
finitely supported on the following $t$-dependent region of the
cone of dominant weights $\mathcal{P}^+$
\begin{equation}
\mathcal{P}^{+}_{\text{clas}}(t)=
\begin{cases}
\{ \lambda\in \mathcal{P}^+ \mid \rho+\lambda\in
t\hat{w}(\mathbf{V}_{\text{clas}}) \}  &\text{for}\; t> 0 ,\\
\{ \lambda\in \mathcal{P}^+ \mid \rho+\lambda\in
tw_0\hat{w}(\mathbf{V}_{\text{clas}}) \}  &\text{for}\; t< 0 .
\end{cases}
\end{equation}
Because of dimensional considerations, it is clear that the
cardinality of the support $\mathcal{P}^{+}_{\text{clas}}(t)$
grows at most polynomially in $t$
\begin{equation}
|\mathcal{P}^{+}_{\text{clas}}(t)|=O(t^N)\qquad\text{for}\;\;
|t|\rightarrow\infty .
\end{equation}
The classical wave packet is defined as
\begin{eqnarray}\label{clas:wp}
\lefteqn{\phi^{(\text{clas})}_{\lambda} (t) =} && \\
&& \begin{cases} {\displaystyle
\frac{(-1)^{\hat{w}}}{|W|\text{Vol}(\mathbf{A})} \int_{\mathbf{A}}
e^{i\langle \rho+\lambda
,\xi_{\hat{w}}\rangle-it\hat{E}(\xi)}\hat{\phi}(\xi) \text{d}\xi }
&\text{for}\;\lambda\in
\mathcal{P}^{+}_{\text{clas}}(t)\;\text{and}\; t>0 ,
\\[3ex]
{\displaystyle \frac{(-1)^{w_0\hat{w}}}{|W|\text{Vol}(\mathbf{A})}
\int_{\mathbf{A}} e^{i\langle \rho+\lambda
,\xi_{w_0\hat{w}}\rangle-it\hat{E}(\xi)}\hat{\phi}(\xi)
\text{d}\xi } &\text{for}\;\lambda\in
\mathcal{P}^{+}_{\text{clas}}(t)\;\text{and}\; t<0 ,\\[0.5ex]
\makebox[6em]{} 0 & \text{otherwise}.
\end{cases} \nonumber
\end{eqnarray}

The next lemma compares the long-time asymptotics of the free wave
packet $\phi^{(0)}(t)$ with that of the classical wave packet
$\phi^{(\text{clas})}(t)$.
\begin{lemma}\label{clas:lem}
For $t\to\pm \infty$, the difference between the free wave packet
$\phi^{(0 )}(t)$ and the classical wave packet
$\phi^{(\text{clas})}(t)$ tends to zero:
\begin{equation*}
\forall\kappa >0:\qquad  \|
\phi^{(0)}(t)-\phi^{(\text{clas})}(t)\|_{\mathcal{H}} =O(1/|t|^\kappa)
\quad\text{as}\;t\to\pm\infty .
\end{equation*}
\end{lemma}
\begin{proof}
It is immediate from the definitions of the wave packets under
consideration that
\begin{subequations}
\begin{equation}\label{qwavea}
\phi^{(0)}_\lambda(t)-\phi^{(\text{clas})}_\lambda(t) =
\frac{1}{|W|\,\text{Vol} (\mathbf{A})} \sum_{w\in \hat{W}}
(-1)^w\, \int_{\mathbf{A}} e^{i\langle \rho+\lambda ,
\xi_w\rangle-it\hat{E}(\xi) } \hat{\phi} (\xi) \text{d}\xi ,
\end{equation}
with
\begin{equation}\label{qwaveb}
\hat{W}\equiv
\begin{cases}
W\setminus \{ \hat{w}\} & \text{if}\;
\lambda\in\mathcal{P}^+_{\text{clas}}(t)\;\text{and}\;t>0 ,\\
W\setminus \{ w_0\hat{w}\} & \text{if}\;
\lambda\in\mathcal{P}^+_{\text{clas}}(t)\;\text{and}\;t<0 ,\\
W & \text{otherwise}.
\end{cases}
\end{equation}
\end{subequations}
The proof of the lemma now hinges on a stationary phase estimate
extracted from the Corollary of Theorem XI.14 in Ref. \cite[p.
38-39]{ree-sim:methods}, which states that for any $k>0$ there
exists a (positive) constant $c_k$ such that
\begin{subequations}
\begin{equation}\label{spa}
\left| \int_{\mathbf{A}} e^{i\langle \mathbf{x},\xi\rangle -it
\hat{E}(\xi)} \hat{\phi}(\xi) \text{d}\xi \right|\leq
\frac{c_k}{(1+|\mathbf{x}|+|t|)^k}
\end{equation}
for all $\mathbf{x}\in\mathbf{E}$ and $t\in\mathbb{R}$ such that
\begin{equation}\label{spb}
\mathbf{x}\not\in t \mathbf{V}_{\text{clas}}.
\end{equation}
\end{subequations}
Indeed, invoking of the stationary phase estimate in Eqs.
\eqref{spa}, \eqref{spb} with $k> N/2$ and $\mathbf{x}=w^{-1}
(\rho+\lambda)$, reveals that the norm of the
difference between the wave packets given
by Eqs. \eqref{qwavea}, \eqref{qwaveb} in the
Hilbert space $\mathcal{H}$ is $O(1/|t|^{k-N/2})$ as
$t\to\pm\infty$. (Notice in this
connection that $w^{-1} (\rho+\lambda)\in
t\mathbf{V}_{\text{clas}}$ if and only if
$\lambda\in\mathcal{P}^+_{\text{clas}}(t)$ and $w\in
W\setminus\hat{W}$.)
\end{proof}

\subsection{The Asymptotic Wave Packet}\label{sec5.3}
Let us define {\em asymptotic wave packets}
$\phi^{(\infty)}_{\pm}(t)$ that are obtained from the interacting
wave packets $\phi_{\pm}(t)$ upon replacing the Fourier kernel
$\Psi_\lambda (\xi)$ by its plane wave asymptotics
$\Psi_\lambda^{(\infty )} (\xi)$
\begin{subequations}
\begin{equation}\label{as:wp}
\phi^{(\infty)}_{\pm,\lambda }(t) =
\frac{1}{|W|\text{Vol}(\mathbf{A})} \int_{\mathbf{A}}
\Psi_\lambda^{(\infty )} (\xi) e^{-it\hat{E}(\xi)}
\hat{\mathcal{S}}_L^{\mp 1/2} (\xi) \hat{\phi}(\xi) \text{d}\xi ,
\end{equation}
or more explicitly
\begin{eqnarray}
\phi^{(\infty)}_{+,\lambda }(t)\!\! &=&
\!\!\frac{1}{|W|\,\text{Vol} (\mathbf{A})} \sum_{w\in W} (-1)^w\,
\int_{\mathbf{A}} e^{i\langle \rho+\lambda ,
\xi_w\rangle-it\hat{E}(\xi) }
\frac{\hat{\mathcal{C}}(\xi_w)}{\hat{\mathcal{C}}(\xi_{\hat{w}})}
\hat{\phi} (\xi)\text{d}\xi , \\
\phi^{(\infty)}_{-,\lambda }(t)\!\! &=&\!\!
\frac{1}{|W|\,\text{Vol} (\mathbf{A})} \sum_{w\in W} (-1)^w\,
\int_{\mathbf{A}} e^{i\langle \rho+\lambda ,
\xi_w\rangle-it\hat{E}(\xi) }
\frac{\hat{\mathcal{C}}(\xi_w)}{\hat{\mathcal{C}}(\xi_{w_0\hat{w}})}
\hat{\phi} (\xi)\text{d}\xi .
\end{eqnarray}
\end{subequations}
The next lemma states that the long-time behavior of the
asymptotic wave packets is governed by the classical wave packet
$\phi^{(\text{clas})}(t)$ \eqref{clas:wp}.

\begin{lemma}\label{as1:lem}
For $t\to\pm \infty$, the difference between the asymptotic wave
packet $\phi^{(\infty )}_\pm (t)$ and the classical wave packet
$\phi^{(\text{clas})}(t)$ tends to zero:
\begin{equation*}
\forall \kappa >0:\qquad \| \phi^{(\infty)}_\pm
(t)-\phi^{(\text{clas})}(t)\|_{\mathcal{H}} =O(1/|t|^\kappa)
\quad\text{as}\;t\to\pm\infty .
\end{equation*}
\end{lemma}
\begin{proof}
The proof of Lemma \ref{clas:lem} applies verbatim, upon replacing
$\phi^{(0)}(t)$ by $\phi^{(\infty )}_\pm (t)$ and the introduction
of minor modifications in the formulas so as to incorporate the
additional (smooth) factors
$\hat{\mathcal{C}}(\xi_w)/\hat{\mathcal{C}}(\xi_{\hat{w}})$ and
$\hat{\mathcal{C}}(\xi_w)/\hat{\mathcal{C}}(\xi_{w_0\hat{w}})$,
respectively.
\end{proof}

Let $P^{(\text{clas})}_t:\mathcal{H}\mapsto\mathcal{H}$ denote the
orthogonal projection onto the finite-dimensional subspace
$l^2(\mathcal{P}^+_{\text{clas}}(t))\subset l^2(\mathcal{P}^+)$:
\begin{equation}
(P^{(\text{clas})}_t\phi)_\lambda =
\begin{cases}
\phi_\lambda & \text{if}\; \lambda\in\mathcal{P}^+_{\text{clas}}(t) , \\
0 & \text{if}\;
\lambda\in\mathcal{P}^+\setminus\mathcal{P}^+_{\text{clas}}(t).
\end{cases}
\end{equation}
It is clear from the definition of the classical wave packet that
$P^{(\text{clas})}_t(\phi^{(\text{clas})}(t))=\phi^{(\text{clas})}(t)$.
As a consequence, we get from from Lemma \ref{as1:lem} upon
projection onto $l^2(\mathcal{P}^+_{\text{clas}}(t))$ that
\begin{equation}\label{as1:eq}
\forall \kappa >0:\qquad \| P^{(\text{clas})}_t\phi^{(\infty)}_\pm
(t)-\phi^{(\text{clas})}(t)\|_{\mathcal{H}} = O(1/|t|^\kappa)\quad
\text{as}\; t\to\pm\infty.
\end{equation}

Our final lemma states that the long-time asymptotics of the
interacting wave packet $\phi_{\pm}(t)$ coincides with that of the
asymptotic wave packet $\phi^{(\infty)}_\pm (t)$.

\begin{lemma}\label{as2:lem}
For $t\to\pm \infty$, the difference between the interacting wave
packet $\phi_{\pm }(t)$ and the asymptotic wave packet
$\phi^{(\infty )}_\pm (t)$ tends to zero:
\begin{equation*}
\forall \kappa >0:\qquad \| \phi_{\pm}(t)-\phi^{(\infty )}_\pm
(t)\|_{\mathcal{H}} = O(1/|t|^\kappa)\quad\text{as}\ t\to\pm\infty
.
\end{equation*}
\end{lemma}

\begin{proof}
From the definitions it is immediate that
\begin{equation*}
\phi_{\pm,\lambda}(t)-\phi^{(\infty )}_{\pm,\lambda}(t) =
(e^{-it\hat{E}}\hat{\mathcal{S}}_L^{\mp 1/2} \hat{\phi},
\overline{\Psi}_\lambda-
\overline{\Psi}_\lambda^{(\infty)})_{\hat{\mathcal{H}}}.
\end{equation*}
Hence
\begin{eqnarray*}
\| P_t^{(\text{clas})}\left( \phi_{\pm}(t)-\phi^{(\infty )}_\pm
(t)\right) \|_{\mathcal{H}}^2&=&
\sum_{\lambda\in\mathcal{P}^+_{\text{clas}}(t)} |
(e^{-it\hat{E}}\hat{\mathcal{S}}_L^{\mp 1/2} \hat{\phi},
\overline{\Psi}_\lambda-\overline{\Psi}_\lambda^{(\infty)})_{\hat{\mathcal{H}}}|^2
\\
&\leq & \| \hat{\phi}\|_{\hat{\mathcal{H}}}^2
\sum_{\lambda\in\mathcal{P}^+_{\text{clas}}(t)}
\|\Psi_\lambda-\Psi_\lambda^{(\infty)}\|_{\hat{\mathcal{H}}}^2
\end{eqnarray*}
(by the Cauchy-Schwarz inequality). Now, since $|
\mathcal{P}^+_{\text{clas}}(t)|=O(t^N)$ and $m(\lambda)
>|t|\varepsilon -1$ for $\lambda\in \mathcal{P}^+_{\text{clas}}(t)$,
we conclude from this estimate combined with Theorem
\ref{plane:thm} that $\| P^{(\text{clas})}_t\left(
\phi_{\pm}(t)-\phi^{(\infty )}_\pm (t)\right) \|_{\mathcal{H}}$
converges to zero exponentially fast for $t\to\pm\infty$, so in particular
\begin{equation}\label{as2:eq}
\forall\kappa >0:\qquad \| P^{(\text{clas})}_t\left(
\phi_{\pm}(t)-\phi^{(\infty )}_\pm (t)\right) \|_{\mathcal{H}} =
O(1/|t|^\kappa)\quad\text{as}\; t\to\pm\infty.
\end{equation}
The lemma now follows from the vanishing of the tails
$(\text{Id}-P^{(\text{clas})}_t)\phi^{(\infty )}_\pm (t)$ and
$(\text{Id}-P^{(\text{clas})}_t) \phi_{\pm }(t)$ for
$t\to\pm\infty$:
\begin{subequations}
\begin{eqnarray}
&&  \|
(\text{Id}-P^{(\text{clas})}_t)\phi^{(\infty
)}_\pm (t) \|_{\mathcal{H}} =O(1/|t|^\kappa)\quad\text{as}\;t\to\pm\infty
\label{tail1} ,\\
&& \| (\text{Id}-P^{(\text{clas})}_t)
\phi_{\pm }(t) \|_{\mathcal{H}} =O(1/|t|^\kappa)\quad\text{as}\; t\to\pm\infty .
\label{tail2}
\end{eqnarray}
\end{subequations}
Notice in this connection that the limiting relation in Eq.
\eqref{tail1} is immediate from Lemma \ref{as1:lem} and Eq.
\eqref{as1:eq}, and that the limiting relation in Eq.
\eqref{tail2} follows by compairing the norm equality
\begin{equation*}
\| \phi_{\pm}(t)\|_{\mathcal{H}}=\| \hat{\phi}
\|_{\hat{\mathcal{H}}}
\end{equation*}
with the norm estimate for $t\to\pm\infty$
\begin{eqnarray*}
\| P^{(\text{clas})}_t\phi_{\pm }(t)
\|_{\mathcal{H}} &\stackrel{\text{Eq.}~\eqref{as2:eq}}{=}&
\|
P^{(\text{clas})}_t\phi^{(\infty )}_\pm (t) \|_{\mathcal{H}}
+ O(1/|t|^\kappa) \\
&\stackrel{\text{Eq.}~\eqref{as1:eq}}{=}&  \| \phi^{(clas )}_t \|_{\mathcal{H}}
+ O(1/|t|^\kappa) \\
&\stackrel{\text{Lemma}~\ref{clas:lem}}{=}&
\|
\phi^{(0 )}(t) \|_{\mathcal{H}}+ O(1/|t|^\kappa) \\
&=& \| \hat{\phi} \|_{\hat{\mathcal{H}}}+ O(1/|t|^\kappa) ,
\end{eqnarray*}
which entails that
\begin{equation*}
\| (\text{Id}-P^{(\text{clas})}_t)
\phi_{\pm }(t) \|_{\mathcal{H}}=
\sqrt{\| \phi_{\pm}(t)\|_{\mathcal{H}}^2-\| P^{(\text{clas})}_t\phi_{\pm }(t)
\|_{\mathcal{H}}^2}=O(1/|t|^{\kappa /2}).
\end{equation*}
\end{proof}

\subsection{Proof of Proposition \ref{asf:prp}}\label{sec5.4}
After these preparations, the proof of Proposition \ref{asf:prp}
reduces to the telescope
\begin{eqnarray}
\lefteqn{\| \phi_{\pm}(t)-\phi^{(0)}(t)\|_{\mathcal{H}}} &&
\\&& \leq \| \phi_{\pm}(t)-\phi^{(\infty )}_\pm (t)\|_{\mathcal{H}} + \|
\phi^{(\infty )}_\pm (t)-\phi^{(clas )}(t)\|_{\mathcal{H}} + \|
\phi^{(clas)}(t)-\phi^{(0 )}(t)\|_{\mathcal{H}}, \nonumber
\end{eqnarray}
and the application of Lemmas \ref{clas:lem}, \ref{as1:lem} and
\ref{as2:lem}.

\section{Application: Scattering of Hyperbolic Lattice Calogero-Moser
Models}\label{sec6}

In this section we specialize our $c$-functions so as to describe
the scattering of the hyperbolic Ruijsenaars-Schneider type
lattice Calogero-Moser models associated with the Macdonald
polynomials. Initially, viz. in the first three subsections, it
will be assumed that our root system $\boldsymbol{R}$ be {\em
reduced} (so $\boldsymbol{R}_0=\boldsymbol{R}_1=\boldsymbol{R}$)
except when explicitly stated otherwise. In the fourth subsection,
we then indicate how the results extend to the case of a {\em
nonreduced} root system (so $\boldsymbol{R}=BC_N$,
$\boldsymbol{R}_0=C_N$ and $\boldsymbol{R}_1=B_N$). We end our
study of the lattice Calogero-Moser models in the fifth subsection
by providing some illuminating additional details describing what
the results boil down to in the simplest situation of a root
system of rank {\em one}.

\subsection{Macdonald Wave Function}\label{sub51}\label{sec6.1}
For $c$-functions of the form
\begin{equation}\label{m-c}
\hat{c}_{|\alpha|}(z) =
\frac{(q^{g_{|\alpha|}}z;q)_\infty}{(qz;q)_\infty},\qquad
q=e^{-s}, \quad g_{|\alpha|},s >0,
\end{equation}
with $(z;q)_\infty\equiv\prod_{n=0}^\infty (1-zq^n)$, the weight
function $\hat{\Delta} (\xi)$ \eqref{p-m}--\eqref{c-f} becomes
\begin{equation}\label{m-weight}
\hat{\Delta} (\xi)= \prod_{\alpha\in \boldsymbol{R}}
\frac{(qe^{i\langle \alpha ,\xi\rangle};q)_\infty}
      {(q^{g_{|\alpha|}} e^{i\langle \alpha
      ,\xi\rangle};q)_\infty}.
\end{equation}
(The positivity restrictions on the parameters $g_{|\alpha|}$ and
$s$ guarantee that the $c$-function \eqref{m-c} meets the
technical requirements stated in Section \ref{sec2}.) Our
polynomials $P_\lambda (\xi)$, $\lambda\in\mathcal{P}^+$ now
amount to the orthonormalized Macdonald polynomials
\cite{mac:symmetric,mac:orthogonal,mac:affine}
\begin{subequations}
\begin{equation}\label{omp}
P_\lambda (\xi) =\frac{1}{\mathcal{N}_0^{1/2}}
\Delta^{1/2}(\lambda) \mathbf{P}_\lambda (\xi) ,
\end{equation}
where
\begin{equation}\label{hm}
 \Delta (\lambda)= \frac{\mathcal{C}^+
(\rho_g)\mathcal{C}^- (\rho_g)}{\mathcal{C}^+
(\rho_g+\lambda)\mathcal{C}^- (\rho_g+\lambda)},\qquad
\mathcal{N}_0 =\frac{\mathcal{C}^- (\rho_g)}{\mathcal{C}^+
(\rho_g)},
\end{equation}
with $\rho_{g} \equiv \frac{1}{2}\sum_{\alpha\in \boldsymbol{R}^+}
g_{|\alpha |}\,\alpha$ and
\begin{eqnarray}\label{cm-f} \mathcal{C}^\pm
(\mathbf{x})&=&\prod_{\alpha\in \boldsymbol{R}^+} c^\pm_{|\alpha
|}(\langle\mathbf{x},\alpha^\vee \rangle ) , \\
c^+_{|\alpha |}(x) &=&  q^{g_{|\alpha|} x /2}
\frac{(q^{g_{|\alpha|}+x};q)_\infty}
     {(q^{x};q)_\infty} , \label{cm-f1} \\
c^-_{|\alpha |}(x) &=&  q^{g_{|\alpha|} x/2} \frac{(
q^{1+x};q)_\infty}
     {(q^{1-g_{|\alpha|}+x};q)_\infty}. \label{cm-f2}
\end{eqnarray}
\end{subequations}
 Here $\mathbf{P}_\lambda (\xi )$ denotes the
Macdonald polynomial
\begin{subequations}
\begin{equation}\label{mp}
\mathbf{P}_\lambda (\xi )= c_\lambda p_\lambda (\xi),\qquad
c_\lambda = \frac{\mathcal{C}^+(\rho_g+\lambda
)}{\mathcal{C}^+(\rho_g )} ,
\end{equation}
where
\begin{equation}\label{mp1}
p_\lambda (\xi) =  m_\lambda (\xi) + \sum_{\mu\in\mathcal{P}^+,\,
\mu \prec \lambda} c_{\lambda\mu} m_\mu (\xi),\qquad
\end{equation}
with coefficients $c_{\lambda\mu}\in\mathbb{C}$ such that
\begin{equation}\label{mp2}
( p_\lambda , m_\mu )_{\hat{\Delta}} = 0  \quad \text{for}\; \mu
\prec\lambda .
\end{equation}
\end{subequations}
For the Macdonald weight function $\hat{\Delta}(\xi)$
\eqref{m-weight}, the coefficients $c_{\lambda\mu}$ turn out to
vanish when $\lambda$ and $\mu$ are not comparable in the
dominance ordering \cite{mac:orthogonal}. In other words, in this
case one may take the ordering $\succeq$ to be equal to the
dominance order $\geqslant$ \eqref{po} without restricting
generality. Explicit formulas for the expansion coefficients
$c_{\lambda\mu}$ when $\mathcal{P}\neq\mathcal{Q}$ (i.e. excluding
the root systems $E_8$, $F_4$ and $G_2$) can be found in Ref.
\cite{die-lap-mor:determinantal}.

We thus arrive at the following formula for the wave function
$\Psi_\lambda (\xi) $ \eqref{wave-f} in terms of Macdonald
polynomials.

\begin{proposition}[Macdonald Wave Function]\label{mwf:prp}
For $\boldsymbol{R}$ reduced and $c$-functions given by
$\hat{c}_{|\alpha|}(z)$ \eqref{m-c}, the wave function
$\Psi_\lambda (\xi) $ \eqref{wave-f} reads explicitly
\begin{equation*}
\Psi_\lambda (\xi) = \frac{1}{\mathcal{N}_0^{1/2}}\Delta^{1/2}
(\lambda)\hat{\Delta}^{1/2}(\xi) \delta (\xi) \mathbf{P}_\lambda
(\xi),
\end{equation*}
with  $\mathbf{P}_\lambda (\xi)$ denoting the Macdonald polynomial
characterized by Eqs. \eqref{mp}--\eqref{mp2}.
\end{proposition}

\subsection{Macdonald-Ruijsenaars Laplacian}\label{sec6.2}
Let us recall that a nonzero weight $\pi\in\mathcal{P}^+$ is called {\em
minuscule} if $\langle \pi,\alpha^\vee\rangle \leq 1$ for all
$\alpha\in \boldsymbol{R}^+$ and that it is called {\em
quasi-minuscule} if $\langle \pi,\alpha^\vee\rangle \leq 1$ for
all $\alpha\in \boldsymbol{R}^+\setminus \{ \pi\}$ (and it is not
minuscule). The number of minuscule weights is equal to the index
of $\mathcal{Q}$ in $\mathcal{P}$ minus $1$ (so for $E_8$, $F_4$ and $G_2$
there are none). As regards to the quasi-minuscule weights: there
is always just {\em one} and it is given by $\pi=\alpha_0$, where
$\alpha_0^\vee$ is the maximal root of the dual root system
$\boldsymbol{R}^\vee\equiv \{\alpha^\vee\mid\alpha\in\boldsymbol{R}\}$. For
the readers convenience, we have included a list of the
(quasi-)minuscule weights for each root system in Table
\ref{qmweight} (where we have adopted the standard numbering of
the fundamental weights in accordance with Refs.
\cite{bou:groupes,hum:introduction}).
\begin{table}
\begin{equation*}
\begin{array}{lll}
\boldsymbol{R} & \text{minuscule} & \text{quasi-minuscule} \\[1ex]
A_N :& \omega_1,\ldots ,\omega_{N} & \omega_1+\omega_{N},\\
B_N :& \omega_N  & \omega_1 ,\\
C_N :& \omega_1 & \omega_2 ,\\
D_N :& \omega_1,\omega_{N-1},\omega_N & \omega_2, \\ [1ex]
E_6 :& \omega_1,\omega_6 & \omega_2,\\
E_7 :& \omega_7 & \omega_1 ,\\
E_8: &  &  \omega_8,\\
F_4: &  &  \omega_4,\\
G_2: &  &  \omega_1 ,\\[1ex]
BC_N: &  &  \omega_1 .\\
 &   &
\end{array}
\end{equation*}
\caption{Minuscule and Quasi-Minuscule Weights.}\label{qmweight}
\end{table}

To a (quasi-)minuscule weight $\pi$ we associate the
multiplication operator $\hat{E}_\pi
:\hat{\mathcal{H}}\mapsto\hat{\mathcal{H}}$ given by
\begin{equation}
\hat{E}_\pi (\xi) =\sum_{\nu\in W (\pi )\cup W(-\pi)} \exp
(i\langle \nu , \xi \rangle ).
\end{equation}
The {\em Macdonald-Ruijsenaars Laplacian} is now defined as the
pullback $L_\pi:\mathcal{H}\mapsto\mathcal{H}$ of $\hat{E}_\pi$
with respect to the Fourier transform $\mathcal{F}$
\begin{equation}
L_\pi= \mathcal{F}^{-1}\circ \hat{E}_\pi \circ\mathcal{F} .
\end{equation}
By Proposition \ref{dl:prp}, the Macdonald-Ruijsenaars Laplacian
$L_\pi\in \mathbb{R}[L_1,\ldots ,L_N]$ constitutes a difference
operator in $\mathcal{H}$. The following proposition provides its
explicit action on lattice functions over the dominant cone
$\mathcal{P}^+$.

\begin{proposition}[Macdonald-Ruijsenaars
Laplacian]\label{mrlap:prp} For $\boldsymbol{R}$ reduced and $\pi$
(quasi-)minuscule, the action of the Macdonald-Ruijsenaars
Laplacian $L_\pi$ on a (square-summable) lattice function
$\phi:\mathcal{P}^+\rightarrow \mathbb{C}$ is given by
\begin{eqnarray*}
\lefteqn{L_\pi \phi_\lambda =
E_\pi (\rho_g^\vee)\, \phi_{\lambda} +} && \\
&&  \sum_{\begin{subarray}{c} \nu\in W(\pi )\cup W(-\pi)\\
\lambda+\nu\in\mathcal{P}^+\end{subarray}} \Bigl( V_\nu^{1/2}
(\rho_g +\lambda ) V_{-\nu}^{1/2} (\rho_g +\lambda +\nu
) \phi_{\lambda+\nu} - V_\nu (\rho_g +\lambda )\phi_{\lambda} \Bigr) ,\\
\end{eqnarray*}
where
\begin{eqnarray*}
V_\nu(\mathbf{x})&=& \prod_{\begin{subarray}{c} \alpha\in \boldsymbol{R} \\
\langle \nu ,\alpha^\vee\rangle >0 \end{subarray}}
\frac{(g_{|\alpha |}+\langle \mathbf{x},\alpha^\vee \rangle :
\sinh_s)_{\langle \nu ,\alpha^\vee\rangle}}{(\langle
\mathbf{x},\alpha^\vee \rangle:\sinh_s )_{\langle \nu
,\alpha^\vee\rangle}}
\\
&=&
\prod_{\begin{subarray}{c} \alpha\in \boldsymbol{R} \\
\langle \nu ,\alpha^\vee\rangle =1 \end{subarray}}
\frac{\sinh\frac{s}{2}(g_{|\alpha |}+\langle
\mathbf{x},\alpha^\vee \rangle
)}{\sinh\frac{s}{2} (\langle \mathbf{x},\alpha^\vee \rangle)} \times\\
&&
\prod_{\begin{subarray}{c} \alpha\in \boldsymbol{R} \\
\langle \nu ,\alpha^\vee\rangle =2 \end{subarray}}
\frac{\sinh\frac{s}{2}(g_{|\alpha |}+\langle
\mathbf{x},\alpha^\vee \rangle )}{\sinh\frac{s}{2} (\langle
\mathbf{x},\alpha^\vee \rangle)} \frac{\sinh\frac{s}{2}(1+
g_{|\alpha |}+\langle \mathbf{x},\alpha^\vee \rangle
)}{\sinh\frac{s}{2} (1+\langle \mathbf{x},\alpha^\vee \rangle)} ,
\\
E_\pi (\mathbf{x}) &=&\sum_{\nu\in W (\pi )\cup W(-\pi)} \exp (
s\langle \nu , \mathbf{x} \rangle ) ,
\end{eqnarray*}
with $(z:\sinh_s )_m\equiv \prod_{\ell =0}^{m-1}\sinh
\frac{s}{2}(z+\ell )$ and $\rho_g^\vee \equiv
\frac{1}{2}\sum_{\alpha\in \boldsymbol{R}^+} g_{|\alpha |}
\alpha^\vee$.
\end{proposition}

\begin{proof}
It is a straightforward consequence of the definitions that the
$c$-functions satisfy the difference equations
\begin{equation*}
\frac{c^+_{|\alpha|}(x+1)}{c^+_{|\alpha|}(x)}= \frac{\sinh
(\frac{sx}{2})}{\sinh \frac{s}{2}(g_{|\alpha |} +x)}, \qquad
\frac{c^-_{|\alpha|}(x+1)}{c^-_{|\alpha|}(x)}= \frac{\sinh
\frac{s}{2}(1+x-g_{|\alpha|})}{\sinh \frac{s}{2}(1 +x)} .
\end{equation*}
With the aid of these difference equations it is not difficult to
verify the fundamental functional relation
\begin{equation}\label{ffe}
\Delta (\mathbf{x}+\nu)V_{-\nu}(\rho_g+\mathbf{x} +\nu) =\Delta
(\mathbf{x}) V_\nu(\rho_g+\mathbf{x} ) ,\qquad \nu\in W(\pi).
\end{equation}
From the recurrence relation for the Macdonald polynomials
exhibited in Eq. \eqref{rec-rela} of the Appendix, it is now
readily inferred---upon invoking the functional relation
\eqref{ffe} specialized to $\mathbf{x}=\lambda$ with $\lambda$ and
$\lambda+\nu$ dominant---that the Macdonald wave function
$\Psi_\lambda (\xi )$ \eqref{cm-f}--\eqref{cm-f2} satisfies the
eigenvalue equation
\begin{eqnarray*}
\sum_{\begin{subarray}{c}\nu\in W(\pi )\\
\lambda+\nu\in\mathcal{P}^+\end{subarray}} \Bigl( V_\nu^{1/2}
(\rho_g +\lambda ) V_{-\nu}^{1/2} (\rho_g +\lambda +\nu )
\Psi_{\lambda+\nu} (\xi ) -V_\nu (\rho_g +\lambda )
\Psi_{\lambda}(\xi) \Bigr)  && \\
=\sum_{\nu\in W(\pi)} \bigl( e^{i\langle \nu ,\xi\rangle}
-q^{\langle \nu \rho_g^\vee\rangle} \bigr) \Psi_{\lambda}(\xi) .
&&
\end{eqnarray*}
Combining with the corresponding eigenvalue equation in which
$\pi$ is replaced by $-w_0(\pi)$ (where $w_0$ is the longest
element of the Weyl group), leads us to the eigenvalue equation
for $L_\pi$:
\begin{eqnarray*}
\sum_{\begin{subarray}{c} \nu\in W(\pi )\cup W(-\pi)\\ \lambda
+\nu\in\mathcal{P}^+
\end{subarray}} \bigl( V_\nu^{1/2} (\rho_g +\lambda ) V_{-\nu}^{1/2}
(\rho_g +\lambda +\nu ) \Psi_{\lambda+\nu} (\xi ) -V_\nu (\rho_g
+\lambda ) \Psi_{\lambda}(\xi) \Bigr)  &&\\
 =\bigl( \hat{E}_\pi (\xi) -E_\pi (\rho_g^\vee) \bigr) \Psi_{\lambda}(\xi) .
 &&
\end{eqnarray*}
The proposition now follows from the completeness of the Macdonald
wave functions $\Psi_\lambda (\xi )$, $\xi\in\mathbf{A}$ in the
Hilbert space  $\mathcal{H}$.
\end{proof}

When $\pi$ is minuscule we have that
\begin{subequations}
\begin{equation}
V_\nu(\mathbf{x}) = \prod_{\begin{subarray}{c} \alpha\in \boldsymbol{R} \\
\langle \nu ,\alpha^\vee\rangle =1 \end{subarray}}
\frac{\sinh\frac{s}{2}(g_{|\alpha |}+\langle
\mathbf{x},\alpha^\vee \rangle )}{\sinh\frac{s}{2} (\langle
\mathbf{x},\alpha^\vee \rangle)} ,
\end{equation}
and that
\begin{equation}
\sum_{\nu\in W(\pi )\cup W(-\pi)} V_\nu (\mathbf{x} )  =E_\pi
(\rho_g^\vee)
\end{equation}
(by the Macdonald identity in Eq. \eqref{mac-id} of the Appendix).
As a consequence, the action of the Macdonald-Ruijsenaars
Laplacian in Proposition \ref{mrlap:prp} simplifies in this
situation to
\begin{equation}\label{mrlap}
L_\pi \phi_\lambda =  \sum_{\begin{subarray}{c}\nu\in W(\pi )\cup
W(-\pi)\\\lambda+\nu\in\mathcal{P}^+\end{subarray}} V_\nu^{1/2}
(\rho_g +\lambda ) V_{-\nu}^{1/2} (\rho_g +\lambda +\nu )
\phi_{\lambda+\nu} .
\end{equation}
\end{subequations}

For the root system $A_N$, all fundamental weights are minuscule
(cf. Table \ref{qmweight}). Hence, in this special case the
discrete Laplacians $L_1,\ldots ,L_N$ of Section \ref{sec3} are
given by the Macdonald-Ruijsenaars Laplacians $L_{\pi}$
\eqref{mrlap} with $\pi=\omega_j$, $j=1,\ldots ,N$. The operators
in question correspond to the commuting quantum integrals of the
hyperbolic relativistic lattice Calogero-Moser model due to
Ruijsenaars \cite{rui:finite-dimensional,rui:systems}. For the
other root systems, only a small part of the discrete Laplacians
$L_1,\ldots ,L_N$ can be made explicit by means of the
Macdonald-Ruijsenaars Laplacian of Proposition \ref{mrlap:prp} and
Table \ref{qmweight}. In principle, the higher-order commuting
Laplacians may be constructed with the aid of the corresponding
Dunkl-Cherednik difference-reflection operators
\cite{che:double,che:macdonalds}, however, at present explicit
formulas for a set of generators for the algebra of commuting
Laplacians $\mathbb{R}[L_1,\ldots ,L_N]$ are available only in the
case of the {\em classical} root systems
\cite{die:self-dual,sah:nonsymmetric}.

It is of course a consequence of our construction that the algebra
of Laplacians  $\mathbb{R}[L_1,\ldots ,L_N]$ consists of bounded
self-adjoint operators in the Hilbert space $\mathcal{H}$. For the
Macdonald-Ruijsenaars Laplacian $L_\pi$, one can also check this
fact independently directly from the explicit action in Propostion
\ref{mrlap:prp}.

\begin{note}
If $\pi$ is quasi-minuscule then $-\pi\in W(\pi)$ (as $\pi$ is a
root). Thus, in this case $W(-\pi)= W(\pi)$. The same
simplification also occurs for $\pi$ not necessarily
quasi-minuscule when $-\mathbf{1}\in W$ (i.e. when the longest
Weyl-group element $w_0$ equals $-\mathbf{1}$). This {\em is} the
case for the root systems $B_N$, $C_N$, $D_N$ ($N\geq 4$, even),
$E_7$, $E_8$, $F_4$, $G_2$ and $BC_N$, but it is {\em not} the
case for the root systems $A_N$ ($N\geq2$), $D_N$ ($N\geq 3$, odd)
and $E_6$.
\end{note}

\subsection{Scattering Matrix}\label{sub53}\label{sec6.3}
When $g_{|\alpha|}\to 1$, $\forall \alpha\in \boldsymbol{R}$, the
Macdonald $c$-functions $\hat{c}_{|\alpha |}(z)$ \eqref{m-c}
specialize to the unit $c$-function. The Macdonald-Ruijsenaars
Laplacian $L_\pi$ of Proposition \ref{mrlap:prp} reduces in this
limit to the free Laplacian $L_\pi^{(0)}=
(\mathcal{F}^{(0)})^{-1}\circ \hat{E}_\pi \circ\mathcal{F}^{(0)}$
given by
\begin{equation}\label{fl}
L_\pi^{(0)} \phi_\lambda = \sum_{\nu\in W(\pi)\cup W(-\pi)}
\phi_{\lambda +\nu}
\end{equation}
with boundary conditions as stipulated in Proposition
\ref{fl:prp}. The following proposition provides a somewhat more
explicit characterization of these boundary conditions (in the
case of $\pi$ (quasi-)minuscule).

\begin{proposition}[Action of the Free Laplacian]\label{fl-qm:prp}
For $\pi$ (quasi-)minuscule, the action of the free Laplacian
$L_\pi^{(0)}$ is of the form
\begin{equation*}
L_\pi^{(0)} \phi_\lambda =   - n_\pi(\lambda) \phi_\lambda+
\sum_{\begin{subarray}{c} \nu\in W(\pi)\cup W(-\pi) \\
\lambda +\nu \in\mathcal{P}^+\end{subarray}} \phi_{\lambda +\nu} ,
\end{equation*}
with $n_\pi(\lambda)=0$ if $\pi$ is minuscule, and with
$n_\pi(\lambda)$ denoting the number of short simple roots
$\alpha_j$ perpendicular to $\lambda$ if $\pi$ is quasi-minuscule
(where, by convention, all roots qualify as short if
$\boldsymbol{R}$ is simply laced).
\end{proposition}
\begin{proof}
Starting point is the action of $L_\pi^{(0)}$ in Eq. \eqref{fl}
with boundary conditions as detailed in Proposition \ref{fl:prp}.
If $\lambda +\nu \not\in\mathcal{P}^+$, then there exists a simple
root $\alpha_j$ such that $\langle \lambda
+\nu,\alpha_j^\vee\rangle <0$. Hence, since
$\lambda\in\mathcal{P}^+$ and $\nu\in W(\pi)\cup W(-\pi)$ with
$\pi$ (quasi-)minuscule, it follows that we are in either one of
the following three situations:
\begin{itemize}
\item[(i)] $\langle \lambda ,\alpha_j^\vee\rangle =0$ and $\langle
\nu ,\alpha_j^\vee\rangle=-1$,
\item[(ii)] $\langle \lambda ,\alpha_j^\vee\rangle =0$ and $\langle
\nu ,\alpha_j^\vee\rangle=-2$,
\item[(iii)] $\langle \lambda ,\alpha_j^\vee\rangle =1$ and $\langle
\nu ,\alpha_j^\vee\rangle=-2$.
\end{itemize}
It is not difficult to see that in the first and last situation
the weight $\rho+\lambda+\nu$ is stabilized by the simple
reflection $r_{\alpha_j}$. Indeed, we get
\begin{equation*}
r_{\alpha_j}(\rho+\lambda+\nu)=\rho+\lambda+\nu-\langle\rho+\lambda+\nu,\alpha_j^\vee\rangle\,
\alpha_j=\rho+\lambda+\nu
\end{equation*}
(where we exploited the fact that $\langle \rho
,\alpha^\vee\rangle =1$ for $\alpha$ simple). It thus follows that
in these two cases the stabilizer of $\rho+\lambda+\nu$ is
nontrivial, whence the corresponding term $\phi_{\lambda +\nu}$ in
Eq. \eqref{fl} vanishes  by the boundary condition in Proposition
\ref{fl:prp}. The second situation occurs only when $\pi$ is
quasi-minuscule. Clearly we must then have that $\nu = -\alpha_j$,
whence $\langle\rho+\lambda+\nu,\alpha_j^\vee\rangle=-1$ and
$\langle\rho+\lambda+\nu,\alpha_k^\vee\rangle=1+\langle\lambda
,\alpha_k^\vee\rangle-\langle \alpha_j,\alpha_k^\vee\rangle >0$
for $k\neq j$ (where we exploited the fact that $\langle \alpha
,\beta^\vee\rangle \leq 0$ for $\alpha,\beta$ simple and
distinct). It thus follows that the weight $\rho+\lambda+\nu$ is
regular and that the Weyl permutation $w_{\rho+\lambda+\nu}$
taking it to the dominant cone is given by the simple reflection
$r_{\alpha_j}$. Indeed, we now get
\begin{equation*}
r_{\alpha_j}(\rho+\lambda+\nu)=
\rho+\lambda+\nu-\langle\rho+\lambda+\nu,\alpha_j^\vee\rangle\,
\alpha_j=\rho+\lambda .
\end{equation*}
Invoking of the boundary condition in Proposition \ref{fl:prp}
then reveals that the corresponding term $\phi_{\lambda +\nu}$ in
Eq. \eqref{fl} is equal to $-\phi_\lambda$. Now, every simple root
$\alpha_j$ in the Weyl orbit of $\pi$ for which $\langle
\lambda,\alpha_j^\vee\rangle =0$ gives rise to such a contribution
$\phi_{\lambda-\alpha_j}=-\phi_\lambda$ in the action on the
r.h.s. of Eq. \eqref{fl}. Furthermore, since a quasi-minuscule
weight $\pi$ is a short root of $\boldsymbol{R}$ (as
$\pi=\alpha_0$ with $\alpha_0^\vee$ denoting the maximal root of
$\boldsymbol{R}^\vee$, whence $\alpha_0^\vee$ is long and
$\alpha_0$ is short), it is clear that the nonzero contributions
in question occur precisely at all simple short roots
perpendicular to $\lambda$.
\end{proof}

Our main application of the scattering formalism in Section
\ref{sec4} is the following explicit formula for the scattering
and wave operators for the lattice Calogero-Moser system, relating
the long-time asymptotics of the dynamics of the
Macdonald-Ruijsenaars Laplacian $L_\pi$ to that of the free
Laplacian $L_{\pi}^{(0)}$.

\begin{theorem}[Lattice Calogero-Moser Scattering
for $\boldsymbol{R}$ Reduced]\label{lcm:thm} The wave operators
$\Omega_\pm=s-\lim_{t\to \pm\infty} e^{itL_\pi}
e^{-itL_\pi^{(0)}}$ and the scattering operator
$\mathcal{S}_{L_\pi}=\Omega_+^{-1}\Omega_-$ for the
Macdonald-Ruijsenaars Laplacian $L_\pi$ in relation to the free
Laplacian $L_{\pi}^{(0)}$ are of the form stated in Theorem
\ref{wave:thm} and Corollary \ref{scattering:cor}, with a unitary
scattering matrix $\hat{S}_{L_\pi}(\xi)$ given by Eqs. \eqref{Sm},
\eqref{Sw} and
\begin{equation*}
\hat{s}_{|\alpha|}(\langle \alpha , \xi\rangle)= \frac{(q
e^{i\langle \alpha ,\xi\rangle};q)_\infty }{(q^{g_{|\alpha|}}
e^{i\langle \alpha ,\xi\rangle};q)_\infty} \frac{(q^{g_{|\alpha|}}
e^{-i\langle \alpha ,\xi\rangle};q)_\infty }{(q e^{-i\langle
\alpha ,\xi\rangle};q)_\infty } .
\end{equation*}
\end{theorem}

For the type $A$ root systems Theorem \ref{lcm:thm} is due to
Ruijsenaars \cite{rui:factorized}. The scattering of the
corresponding classical-mechanical system was analyzed previously
in Ref. \cite{rui:action-angle}.

\begin{note}
The asymptotics of the Macdonald wave function $\Psi_\lambda
(\xi)$ in Proposition \ref{mwf:prp} is governed by Theorem
\ref{plane:thm} (and Eqs. \eqref{aswave}--\eqref{smat}) with a
scattering matrix taken from Theorem \ref{lcm:thm}.
\end{note}

\subsection{Extension to Nonreduced Root Systems}\label{sec6.4}
We will now indicate how the results of Subsections
\ref{sub51}--\ref{sub53} should be adapted so as to include the
case of a {\em nonreduced} root system (viz.
$\boldsymbol{R}=BC_N$, $\boldsymbol{R}_0= C_N$,
$\boldsymbol{R}_1=B_N$ and $W$ amounts to the hyperoctahedral
group $S_N\ltimes \mathbb{Z}_2^N$). In short, the bottom line is
that all results carry over to the case of nonreduced root systems
upon passing from the Macdonald polynomials to the
Macdonald-Koornwinder multivariate Askey-Wilson polynomials
\cite{koo:askey-wilson,die:self-dual,sah:nonsymmetric}. More
specifically, by picking $c$-functions $\hat{c}_{|\alpha|}(z) $,
$\alpha \in \boldsymbol{R}_1$ of the form
\begin{equation}\label{mk-c}
\hat{c}_{|\alpha|}(z) =
\begin{cases}
{\displaystyle \frac{(q^{\hat{g}}z;q)_\infty}{(qz;q)_\infty}}
&\text{for}\;\alpha\; \text{long},\\[2ex] {\displaystyle
\frac{(q^{\hat{g}_0}z,-q^{\hat{g}_1}z,q^{\hat{g}_2+1/2}z,-q^{\hat{g}_3+1/2}z;q)_\infty}{(qz^2;q)_\infty}}
&\text{for}\;\alpha\; \text{short},
\end{cases}
\end{equation}
where $q=e^{-s}$ and $s, \hat{g},\hat{g}_{0},\ldots ,\hat{g}_3 >0$
(and with $(z_1,\ldots ,z_k;q)_\infty\equiv (z_1;q)_\infty\cdots
(z_k;q)_\infty$), we end up with a weight function $\hat{\Delta}
(\xi)$ \eqref{p-m}--\eqref{c-f} given by
\begin{eqnarray}\label{mk-weight}
\lefteqn{\hat{\Delta} (\xi) = \prod_{\begin{subarray}{c}\alpha\in
\boldsymbol{R}_1\\\alpha\;\text{long}\end{subarray}}
\frac{(qe^{i\langle \alpha ,\xi\rangle};q)_\infty}
      {(q^{\hat{g}} e^{i\langle \alpha
      ,\xi\rangle};q)_\infty} } &&\\
&& \times \prod_{\begin{subarray}{c}\alpha\in
\boldsymbol{R}_1\\\alpha\;\text{short}\end{subarray}} \frac{(qe^{2
i\langle \alpha ,\xi\rangle};q)_\infty}
      {(q^{\hat{g}_0}e^{i\langle \alpha
      ,\xi\rangle},-q^{\hat{g}_1}e^{i\langle \alpha
      ,\xi\rangle},q^{\hat{g}_2+1/2}e^{i\langle \alpha
      ,\xi\rangle},-q^{\hat{g}_3+1/2}e^{i\langle \alpha
      ,\xi\rangle};q)_\infty }. \nonumber
\end{eqnarray}
The polynomials $P_\lambda (\xi)$, $\lambda\in\mathcal{P}^+$
amount in this case to orthonormalized Macdonald-Koornwinder
polynomials
\cite{koo:askey-wilson,die:self-dual,sah:nonsymmetric}. The
polynomials in question are again of the form in Eqs. \eqref{omp},
\eqref{hm} and Eqs. \eqref{mp}--\eqref{mp2}, but now with
\begin{subequations}
\begin{equation}\label{cmk-f}
\mathcal{C}^\pm (\mathbf{x})=\prod_{\alpha\in \boldsymbol{R}^+_1}
c^\pm_{|\alpha |}(\langle\mathbf{x},\alpha \rangle ) ,
\end{equation}
\begin{eqnarray}\label{cmk-f1}
\lefteqn{c^+_{|\alpha |}(x) =} && \\
&& \begin{cases}{\displaystyle  q^{g x/2 } \frac{(q^{g+ x
};q)_\infty}
     {(q^{x};q)_\infty} }&\text{for}\;\alpha\;\text{long} ,\\[2ex]
 {\displaystyle q^{(g_0+g_1+g_2+g_3) x /2}\times} & \\
{\displaystyle
\frac{(q^{g_0+x},-q^{g_1+x},q^{g_2+1/2+x},-q^{g_3+1/2+x};q)_\infty}
     {(q^{2 x};q)_\infty} }&\text{for}\;\alpha\;\text{short},
\end{cases}  \nonumber
\end{eqnarray}
\begin{eqnarray}\label{cmk-f2}
\lefteqn{c^-_{|\alpha |}(x) =} && \\
&&\begin{cases} {\displaystyle q^{g x /2} \frac{(
q^{1+x};q)_\infty}
     {(q^{1-g+x};q)_\infty}} &\text{for}\;\alpha\;\text{long} ,\\[2ex]
{\displaystyle q^{(g_0+g_1+g_2+g_3)x /2}\times }&
\\{\displaystyle \frac{( q^{1+2x};q)_\infty}
     {(q^{1-g_0+x},-q^{1-g_1+x},q^{1/2-g_2+x},-q^{1/2-g_3+x};q)_\infty}}
     &\text{for}\;\alpha\;\text{short},
\end{cases} \nonumber
\end{eqnarray}
\end{subequations}
where we have distinguished dual parameters $g$, $g_0,\ldots ,g_3$
that are related to the parameters $\hat{g}$, $\hat{g}_0,\ldots
,\hat{g}_3$ via the linear relations
\begin{subequations}
\begin{equation}
g=\hat{g},\qquad \left(\begin{array}{c} g_0 \\ g_1 \\g_2 \\ g_3
\end{array}\right)
=\frac{1}{2} \left(\begin{array}{rrrr}
1 & 1 & 1 & 1 \\
1 & 1 & -1 & -1 \\
1 & -1 & 1 & -1 \\
1 & -1 & -1 & 1
\end{array}\right)
\left(\begin{array}{c} \hat{g}_0 \\ \hat{g}_1 \\\hat{g}_2 \\
\hat{g}_3
\end{array}\right) ,
\end{equation}
and with the vectors $\rho_g$ and $\rho_g^\vee$ now taken to be
\begin{eqnarray}
\rho_g &=& \frac{g}{2}\sum_{\begin{subarray}{c} \alpha\in \boldsymbol{R}_1^+\\
\alpha\;\text{long}
\end{subarray}} \alpha +g_0\sum_{\begin{subarray}{c} \alpha\in \boldsymbol{R}_1^+\\
\alpha\;\text{short}
\end{subarray}} \alpha  , \label{rhog1}\\
\rho_g^\vee &=& \frac{\hat{g}}{2}\sum_{\begin{subarray}{c} \alpha\in \boldsymbol{R}_1^+\\
\alpha\;\text{long}
\end{subarray}} \alpha +\hat{g}_0\sum_{\begin{subarray}{c} \alpha\in \boldsymbol{R}_1^+\\
\alpha\;\text{short}
\end{subarray}} \alpha . \label{rhog2}
\end{eqnarray}
\end{subequations}
We thus arrive at the following formula for the wave function
$\Psi_\lambda (\xi)$ \eqref{wave-f} in terms of
Macdonald-Koornwinder polynomials.
\begin{proposition}[Macdonald-Koornwinder Wave
Function]\label{mkwf:prp} For $\boldsymbol{R}$ nonreduced and
$c$-functions given by $\hat{c}_{|\alpha|}(z)$ \eqref{mk-c}, the
wave  function $\Psi_\lambda (\xi) $ \eqref{wave-f} reads
explicitly
\begin{equation*}
\Psi_\lambda (\xi) = \frac{1}{\mathcal{N}_0^{1/2}}\Delta^{1/2}
(\lambda)\hat{\Delta}^{1/2}(\xi) \delta (\xi) \mathbf{P}_\lambda
(\xi),
\end{equation*}
with $\mathbf{P}_\lambda (\xi)$ denoting the Macdonald-Koornwinder
polynomial characterized by Eqs. \eqref{mp}--\eqref{mp2}.
\end{proposition}

From the second-order recurrence relation for the
Macdonald-Koornwinder polynomials \cite{die:self-dual}, we now
obtain the following formula for the action of the
Macdonald-Koornwinder Laplacian $L_\pi= \mathcal{F}^{-1}\circ
\hat{E}_\pi \circ\mathcal{F}$, associated to the first fundamental
weight $\pi=\omega_1$ (which is a quasi-minuscule weight for
$\boldsymbol{R}=BC_N$, cf. Table \ref{qmweight}).

\begin{proposition}[Macdonald-Koornwinder Laplacian]\label{mklap:prp} For $\boldsymbol{R}$
nonreduced and $\pi=\omega_1$, the action of the
Macdonald-Koorwinder Laplacian $L_{\pi}$ on a (square-summable)
lattice function $\phi:\mathcal{P}^+\rightarrow \mathbb{C}$ is
given by
\begin{eqnarray*}
\lefteqn{L_\pi \phi_\lambda =  E_\pi (\rho_g^\vee )\phi_\lambda+} && \\
&& \sum_{\begin{subarray}{c}\nu\in W(\pi )\\
\lambda +\nu\in\mathcal{P}^+\end{subarray}} \Bigl( V_\nu^{1/2}
(\rho_g +\lambda ) V_{-\nu}^{1/2} (\rho_g +\lambda +\nu )
\phi_{\lambda+\nu} -V_\nu (\rho_g +\lambda ) \phi_{\lambda} \Bigr)
,
\end{eqnarray*}
where
\begin{eqnarray*}
V_\nu (\mathbf{x}) &=& \prod_{\begin{subarray}{c} \alpha\in \boldsymbol{R}_1 \\
\alpha \,\text{long},\;\langle \nu ,\alpha\rangle =1
\end{subarray}} \frac{\sinh\frac{s}{2}(g+\langle
\mathbf{x},\alpha \rangle
)}{\sinh\frac{s}{2} (\langle \mathbf{x},\alpha \rangle)} \times\\
&&
 \prod_{\begin{subarray}{c} \alpha\in \boldsymbol{R}_1 \\
\alpha \,\text{short},\;\langle \nu ,\alpha\rangle =1
\end{subarray}}
\frac{\sinh\frac{s}{2}(g_0+\langle \mathbf{x},\alpha \rangle
)}{\sinh\frac{s}{2} (\langle \mathbf{x},\alpha \rangle)}
\frac{\cosh\frac{s}{2}(g_1+\langle \mathbf{x},\alpha \rangle
)}{\cosh\frac{s}{2} (\langle \mathbf{x},\alpha \rangle)} \\
&&
\makebox[5em]{}\times\frac{\sinh\frac{s}{2}(g_2+\frac{1}{2}+\langle
\mathbf{x},\alpha \rangle )}{\sinh\frac{s}{2} (\frac{1}{2}+\langle
\mathbf{x},\alpha \rangle)}
\frac{\cosh\frac{s}{2}(g_3+\frac{1}{2}+\langle \mathbf{x},\alpha
\rangle )}{\cosh\frac{s}{2} (\frac{1}{2}+\langle \mathbf{x},\alpha
\rangle)} , \\
E_\pi (\mathbf{x}) &=& \sum_{\nu\in W(\pi)} \exp (s\langle \nu
,\mathbf{x}\rangle ) ,
\end{eqnarray*}
and with $\rho_g$ and $\rho_g^\vee$ given by Eqs. \eqref{rhog1}
and \eqref{rhog2}, respectively.
\end{proposition}

For $\hat{g}\longrightarrow 1$ and $ \hat{g}_0,\ldots
,\hat{g}_3\longrightarrow 1/2$  (so $g,g_0\longrightarrow 1$ and
$g_1,g_2,g_3\longrightarrow 0$), the $c$-functions
$\hat{c}_{|\alpha |}(z)$ \eqref{mk-c} tend to $1$ (recall in this
connection the duplication formula
$(z^2;q)_\infty=(z,-z,q^{1/2}z,-q^{1/2}z;q)_\infty$). The
Macdonald-Koornwinder Laplacian $L_\pi$ then reduces to the free
Laplacian
\begin{equation}\label{flnr}
L_\pi^{(0)} \phi_\lambda = \sum_{ \nu\in W(\pi),\, \lambda +\nu
\in \mathcal{P}^+} \phi_{\lambda +\nu} .
\end{equation}

Application of the scattering formalism of Section \ref{sec4} now
produces the following scattering and wave operators relating the
long-time asymptotics of the dynamics of the Macdonald-Koornwinder
Laplacian $L_\pi$ to that of the free Laplacian $L_{\pi}^{(0)}$
\eqref{flnr}.

\begin{theorem}[Lattice Calogero-Moser Scattering for $\boldsymbol{R}$
Nonreduced]\label{lcmNR:thm} The wave operators
$\Omega_\pm=s-\lim_{t\to \pm\infty} e^{itL_\pi}
e^{-itL_\pi^{(0)}}$ and the scattering operator
$\mathcal{S}_{L_\pi}=\Omega_+^{-1}\Omega_-$ for the
Macdonald-Koornwinder Laplacian $L_\pi$ in relation to the free
Laplacian $L_{\pi}^{(0)}$ \eqref{flnr} are of the form stated in
Theorem \ref{wave:thm} and Corollary \ref{scattering:cor}, with a
unitary scattering matrix $\hat{\mathcal{S}}_{L_\pi}$ given
by Eqs. \eqref{Sm}, \eqref{Sw} and
\begin{eqnarray*}
\lefteqn{\hat{s}_{|\alpha|}(\langle \alpha , \xi\rangle)=} && \\
&& \begin{cases} {\displaystyle \frac{(q e^{i\langle \alpha
,\xi\rangle};q)_\infty }{(q^{\hat{g}} e^{i\langle \alpha
,\xi\rangle};q)_\infty} \frac{(q^{\hat{g}} e^{-i\langle \alpha
,\xi\rangle};q)_\infty}{(q e^{-i\langle \alpha
,\xi\rangle};q)_\infty } }
 &\text{for}\;\alpha\;\text{long} , \\[2ex]
{\displaystyle \frac{(qe^{2i\langle \alpha ,\xi\rangle}
;q)_\infty}{(q^{\hat{g}_0}e^{i\langle \alpha
,\xi\rangle},-q^{\hat{g}_1}e^{i\langle \alpha
,\xi\rangle},q^{\hat{g}_2+1/2}e^{i\langle \alpha
,\xi\rangle},-q^{\hat{g}_3+1/2}e^{i\langle \alpha
,\xi\rangle};q)_\infty}}\times & \\[2ex]
{\displaystyle \frac{(q^{\hat{g}_0}e^{-i\langle \alpha
,\xi\rangle},-q^{\hat{g}_1}e^{-i\langle \alpha
,\xi\rangle},q^{\hat{g}_2+1/2}e^{-i\langle \alpha
,\xi\rangle},-q^{\hat{g}_3+1/2}e^{-i\langle \alpha ,\xi\rangle}
;q)_\infty}{(qe^{-2i\langle \alpha ,\xi\rangle};q)_\infty}}
 &\text{for}\;\alpha\;\text{short} .
\end{cases}
\end{eqnarray*}
\end{theorem}

\begin{note}
The asymptotics  of the Macdonald-Koornwinder wave function
$\Psi_\lambda (\xi)$ in Proposition \ref{mkwf:prp} is governed by
Theorem \ref{plane:thm} (and Eqs. \eqref{aswave}--\eqref{smat})
with scattering matrices taken from Theorem \ref{lcmNR:thm}.
\end{note}

\subsection{Example: The Rank-One Case}\label{sec6.5}
It is quite instructive to exhibit the results of this section in
somewhat further detail for simplest case of a  root system of
rank {\em one}. We will restrict attention the case of a
nonreduced root system (i.e. $BC_1$), since the reduced case (viz.
$A_1$) can be recovered from it via a specialization of the
parameters (corresponding to a standard reduction from the
Askey-Wilson polynomials to the $q$-ultraspherical polynomials
\cite{ask-wil:some,gas-rah:basic}).

In this situation the Macdonald-Koornwinder wave function takes
the explicit basic hypergeometric form
\begin{equation}
\Psi_l(\xi) = \frac{1}{\mathcal{N}_0^{1/2}} \Delta^{1/2} (\ell )
\hat{\Delta}^{1/2}(\xi ) \delta (\xi)\mathbf{P}_\ell (\xi ),\qquad
\ell\in \mathbb{N}, \quad \xi\in (0,\pi),
\end{equation}
where
\begin{subequations}
\begin{eqnarray}
\mathcal{N}_0 &=& \frac{c^-(g_0)}{c^+(g_0)} ,\\
\Delta (\ell ) &=& \frac{c^+(g_0)c^-(g_0)}{c^+(g_0+\ell)c^-(g_0+\ell)}, \\
\hat{\Delta}(\xi )&=&  \frac{1}{\hat{c}(\xi)\hat{c}(-\xi)}  ,\\
\delta (\xi ) &=&  2\sin(\xi)   ,
\end{eqnarray}
\end{subequations}
with
\begin{subequations}
\begin{eqnarray}
\hat{c}(\xi) &=&
\frac{(q^{\hat{g}_0}e^{-i\xi},-q^{\hat{g}_1}e^{-i\xi},q^{\hat{g}_2+1/2}e^{-i\xi},-q^{\hat{g}_3+1/2}e^{-i\xi};q)_\infty}
{(qe^{-2i\xi};q)_\infty} ,\\
 c^+(x) &=&
 q^{(g_0+g_1+g_2+g_3)  x /2}\times  \\
&&
\frac{(q^{g_0+x},-q^{g_1+x},q^{g_2+1/2+x},-q^{g_3+1/2+x};q)_\infty}
     {(q^{2 x};q)_\infty}   ,  \nonumber \\
 c^-(x) &=&
q^{(g_0+g_1+g_2+g_3)x /2}\times \\
&& \frac{( q^{1+2x};q)_\infty}
     {(q^{1-g_0+x},-q^{1-g_1+x},q^{1/2-g_2+x},-q^{1/2-g_3+x};q)_\infty}, \nonumber
\end{eqnarray}
\end{subequations}
and with $\mathbf{P}_\ell (\xi )$ denoting the Askey-Wilson
polynomial \cite{ask-wil:some,gas-rah:basic}
\begin{equation}
\mathbf{P}_\ell (\xi)=
 {}_4\Phi_3 \left(
\begin{array}{c}
q^{-\ell},q^{2g_0+\ell},q^{\hat{g}_0}e^{i\xi},q^{\hat{g}_0}e^{-i\xi} \\
-q^{g_0+g_1},q^{g_0+g_2+1/2},-q^{g_0+g_3+1/2}
\end{array} ;q,q
\right) .
\end{equation}
Here we have employed standard notation from the theory of basic
hypergeometric series \cite{gas-rah:basic}
\begin{equation*}
{}_{s}\Phi_{s-1} \left( \begin{array}{c} a_1,\ldots ,a_s \\
b_1,\ldots ,b_{s-1}
\end{array} ; q,z \right)\equiv
\sum_{n=0}^\infty \frac{(a_1,\ldots ,a_s ;q)_n }{(b_1,\ldots
,b_{s-1};q)_n } \frac{z^n}{(q;q)_n} ,
\end{equation*}
with $(a;q)_n\equiv\prod_{k=0}^{n-1} (1-aq^k)$ and $(a_1,\ldots
,a_s ;q)_n\equiv(a_1;q)_n\cdots (a_s ;q)_n$.

The asymptotics of the wave function $\Psi_l(\xi)$ for
$\ell\longrightarrow\infty$ is given by (cf. also
\cite{ism-wil:asymptotic,ism:asymptotics})
\begin{subequations}
\begin{equation}
\Psi_l^{\infty}(\xi)=\hat{s}^{1/2}(\xi) e^{i(\ell+1)\xi}-\hat{s}^{-1/2}(\xi)
e^{-i(\ell+1)\xi},
\end{equation}
with
\begin{equation}
\hat{s}(\xi )=\frac{\hat{c}(\xi)}{\hat{c}(-\xi )} .
\end{equation}
\end{subequations}

The free plane waves $\Psi_\ell^{(0)} (\xi )$ \eqref{pwaves} boil
in this case down to the Fourier sine kernel
\begin{equation}
\Psi_\ell^{(0)} (\xi )=2\sin (\ell +1)\xi ,\qquad \ell\in
\mathbb{N}, \quad \xi\in (0,\pi).
\end{equation}

The corresponding Fourier pairings
$\mathcal{F}:l^2(\mathbb{N})\mapsto
L^2((0,\pi),(2\pi)^{-1}\text{d}\xi)$ and
$\mathcal{F}^{(0)}:l^2(\mathbb{N})\mapsto
L^2((0,\pi),(2\pi)^{-1}\text{d}\xi)$ together with their inversion
formulas are given by
\begin{subequations}
\begin{equation}
\begin{cases}
{\displaystyle \hat{\phi}(\xi) = \sum_{\ell\in\mathbb{N}}
\phi_\ell\Psi_\ell(\xi)} , \\[2ex]
{\displaystyle \phi_\ell = \frac{1}{2\pi} \int_0^\pi
\hat{\phi}(\xi)\Psi_\ell(\xi)\text{d}\xi } ,
\end{cases}
\end{equation}
and
\begin{equation}
\begin{cases}
{\displaystyle \hat{\phi}(\xi) = \sum_{\ell\in\mathbb{N}}
\phi_\ell\Psi^{(0)}_\ell(\xi)} , \\[2ex]
{\displaystyle \phi_\ell = \frac{1}{2\pi} \int_0^\pi
\hat{\phi}(\xi)\Psi_\ell^{(0)}(\xi)\text{d}\xi } ,
\end{cases}
\end{equation}
\end{subequations}
respectively (where we have omitted the complex conjugations
because the relevant kernel functions $\Psi_\ell(\xi)$ and
$\Psi_\ell^{(0)}(\xi)$ are real-valued as a consequence of the
fact that $-\mathbf{1}\in W\cong\mathbb{Z}_2$).

The Macdonald-Koornwinder Laplacian $L= \mathcal{F}^{-1}\circ
\hat{E} \circ\mathcal{F}$ and the free Laplacian $L^{(0)}=
(\mathcal{F}^{(0)})^{-1}\circ \hat{E} \circ\mathcal{F}^{(0)}$
associated to the multiplication operator $\hat{E}(\xi) =2\cos
(\xi) $ act on on lattice functions $\phi\in l^2 (\mathbb{N})$
respectively as
\begin{subequations}
\begin{eqnarray}
L \phi_\ell &=& V^{1/2}(g_0+\ell) V^{1/2}(-g_0-\ell-1)\phi_{\ell
+1}+ \\
&& V^{1/2}(-g_0-\ell) V^{1/2}(g_0+\ell-1)\phi_{\ell -1}  + \nonumber\\
&& \bigl( 2\cosh (s\hat{g}_0)-  V(g_0+\ell)
-(1-\delta_{\ell,0})V(-g_0-\ell) \bigr)\phi_\ell ,\nonumber
\end{eqnarray}
with
\begin{eqnarray}
\lefteqn{V(x)=\frac{\sinh\frac{s}{2}(g_0+x )}{\sinh\frac{s}{2}
(x)} \frac{\cosh\frac{s}{2}(g_1+x )}{\cosh\frac{s}{2} (x)} } \\
&& \makebox[2em]{}\times\frac{\sinh\frac{s}{2}(g_2+\frac{1}{2}+x
)}{\sinh\frac{s}{2} (\frac{1}{2}+x)}
\frac{\cosh\frac{s}{2}(g_3+\frac{1}{2}+x )}{\cosh\frac{s}{2}
(\frac{1}{2}+x)} ,\nonumber
\end{eqnarray}
\end{subequations}
and as
\begin{equation}
L^{(0)}\phi_\ell = \phi_{\ell +1}+\phi_{\ell -1} ,
\end{equation}
with the boundary condition $\phi_{-1}=0$.

The specialization of Theorem \ref{lcmNR:thm} to the case $N=1$
now states that the wave operators $\Omega_\pm=s-\lim_{t\to
\pm\infty} e^{itL} e^{-itL^{(0)}}$ and the scattering operator
$\mathcal{S}_{L}=\Omega_+^{-1}\Omega_-$ exist in
$l^2(\mathbb{N})$, and are moreover of the form $\Omega_\pm =
\mathcal{F}^{-1}\circ \hat{\mathcal{S}}^{\mp 1/2}\circ
\mathcal{F}^{(0)}$ and $\mathcal{S}_{L} =
(\mathcal{F}^{(0)})^{-1}\circ \hat{\mathcal{S}}_L\circ
\mathcal{F}^{(0)}$, respectively, with $\hat{\mathcal{S}}_{L}$
being a unitary scattering matrix whose multiplicative action
on a wave packet
$\hat{\phi}\in L^2((0,\pi),(2\pi)^{-1}\text{d}\xi)$ is given
by
\begin{equation}
(\hat{\mathcal{S}}_{L}\hat{\phi})(\xi)=
\frac{\hat{c}(-\xi)}{\hat{c}(\xi
)}\,\hat{\phi}(\xi) \quad\text{for}\; 0 <\xi <\pi .
\end{equation}

\appendix

\section{Properties of the Macdonald Polynomials}\label{appA}
This appendix serves to list a number of key properties of the
Macdonald polynomials $\mathbf{P}_\lambda (\xi)$,
$\lambda\in\mathcal{P}^+$ defined by Eqs. \eqref{mp}--\eqref{mp2}.
We used these properties in Section \ref{sec6} to build the
Macdonald wave function and to the determine the explicit action
of the Macdonald-Ruijsenaars Laplacian. For proofs of the
statements below and for further theory concerning the Macdonald
polynomials the reader is referred to the seminal works of
Macdonald and Cherednik
\cite{mac:symmetric,mac:orthogonal,mac:affine,che:double,che:macdonalds}
(see also \cite{cha:macdonald} for a different approach).
Throughout this appendix it is assumed that our root system
$\boldsymbol{R}$ be {\em reduced}. For the extension of the
statements below to the case of {\em nonreduced} root systems the
reader is referred to Refs.
\cite{mac:orthogonal,mac:affine,koo:askey-wilson,die:self-dual,%
oko:bcn,sah:nonsymmetric}.

The Macdonald polynomials $\mathbf{P}_\lambda (\xi)$
\eqref{mp}--\eqref{mp2} are normalized such that they satisfy the
{\em Specialization Formula}
\begin{equation}\label{sp-f}
\mathbf{P}_\lambda (is\rho_g^\vee) =1.
\end{equation}
In this normalization the {\em Orthogonality Relations} read
\begin{equation}\label{ort-r}
( \mathbf{P}_\lambda , \mathbf{P}_\mu )_{\hat{\mathcal{H}}} =
\begin{cases}
0 &\text{if}\; \lambda \neq \mu ,\\
\frac{\mathcal{N}_0}{\Delta (\lambda )} & \text{if}\; \lambda =\mu
.
\end{cases}
\end{equation}
The specialization formula in Eq. \eqref{sp-f} amounts to the
special case $\mu=0$ of the more general {\em Symmetry Relation}
\begin{equation}\label{sym-r}
\mathbf{P}_\lambda^R (is(\rho_g^\vee +\mu)) =
\mathbf{P}_\mu^{R^\vee} (is(\rho_g +\lambda)),
\end{equation}
where $\mathbf{P}_\lambda^R(\xi) $  and
$\mathbf{P}_\mu^{R^\vee}(\xi)$ refer to the Macdonald polynomials
associated to the root system $\boldsymbol{R}$ and the the dual
root system $\boldsymbol{R}^\vee$, respectively (so $\lambda$ and
$\mu $ are dominant weights of $\boldsymbol{R}$ and
$\boldsymbol{R}^\vee$, respectively).

For any (quasi-)minuscule weight $\pi$ of $\boldsymbol{R}^\vee$,
we have a corresponding {\em Macdonald Difference Equation} given
by
\begin{subequations}
\begin{eqnarray}\label{mdif-eqa}
\sum_{\nu\in W(\pi )} \prod_{\begin{subarray}{c} \alpha\in \boldsymbol{R} \\
\langle \nu ,\alpha\rangle >0 \end{subarray}} \frac{(isg_{|\alpha
|}+\langle \xi ,\alpha \rangle : \sin)_{\langle \nu
,\alpha\rangle}}{(\langle \xi,\alpha \rangle:\sin )_{\langle \nu
,\alpha\rangle}} \bigl(\mathbf{P}_{\lambda
}(\xi+is\nu)-\mathbf{P}_\lambda (\xi) \bigr) &&  \\
  =\sum_{\nu\in W(\pi
)} \bigl( q^{\langle \nu ,\lambda+\rho_g\rangle}-q^{\langle \nu
,\rho_g\rangle}\bigr) \mathbf{P}_\lambda (\xi) ,&& \nonumber
\end{eqnarray}
where $(z:\sin )_m\equiv \prod_{\ell =0}^{m-1}\sin\frac{1}{2}
(z+is\ell )$. If the weight $\pi$ is minuscule (so $|\langle \pi
,\alpha \rangle | \leq 1$, $ \forall \alpha\in \boldsymbol{R}$),
then this difference equation simplifies to
\begin{equation}\label{mdif-eqb}
\sum_{\nu\in W(\pi )} \prod_{\begin{subarray}{c} \alpha\in \boldsymbol{R} \\
\langle \nu ,\alpha\rangle =1 \end{subarray}}
\frac{\sin\frac{1}{2} (isg_{|\alpha |}+\langle \xi ,\alpha \rangle
)}{\sin\frac{1}{2} (\langle \xi,\alpha \rangle )}
\mathbf{P}_{\lambda }(\xi+is\nu)
  =\sum_{\nu\in W(\pi
)} q^{\langle \nu ,\lambda+\rho_g\rangle} \mathbf{P}_\lambda (\xi)
,
\end{equation}
\end{subequations} because of the {\em Macdonald Identity} (for $\pi$ minuscule)
\begin{equation}\label{mac-id}
\sum_{\nu\in W(\pi )}
\prod_{\begin{subarray}{c} \alpha\in \boldsymbol{R} \\
\langle \nu ,\alpha\rangle =1 \end{subarray}}
\frac{\sin\frac{1}{2} (isg_{|\alpha |}+\langle \xi ,\alpha \rangle
)}{\sin\frac{1}{2} (\langle \xi,\alpha \rangle )} = \sum_{\nu\in
W(\pi )} q^{\langle \nu ,\rho_g\rangle} .
\end{equation}

Combination of the Macdonald difference equation in Eq.
\eqref{mdif-eqa} and the symmetry relation in Eq. \eqref{sym-r}
leads to the {\em Recurrence Relation (or Pieri Formula)}
\begin{subequations}
\begin{eqnarray}\label{rec-rela}
\lefteqn{\sum_{\nu\in W(\pi )} \bigl( e^{i\langle \nu
,\xi\rangle}-q^{\langle \nu ,\rho_g^\vee\rangle} \bigr)
\mathbf{P}_\lambda (\xi)=} && \\
&& \sum_{\begin{subarray}{c} \nu\in W(\pi )\\ \lambda
+\nu\in\mathcal{P}^+\end{subarray}} \prod_{\begin{subarray}{c} \alpha\in \boldsymbol{R} \\
\langle \nu ,\alpha^\vee\rangle >0 \end{subarray}}
\frac{(g_{|\alpha |}+\langle \rho_g+\lambda ,\alpha^\vee \rangle :
\sinh_s )_{\langle \nu ,\alpha^\vee\rangle}}{(\langle
\rho_g+\lambda,\alpha^\vee \rangle:\sinh_s )_{\langle \nu
,\alpha^\vee\rangle}} \bigl(\mathbf{P}_{\lambda
+\nu}(\xi)-\mathbf{P}_\lambda (\xi) \bigr) ,\nonumber
\end{eqnarray}
where $\pi$ is now a (quasi-)minuscule weight of $\boldsymbol{R}$
(and $(z:\sinh_s )_m\equiv \prod_{\ell =0}^{m-1}\sinh
(\frac{s}{2}(z+\ell) )$). In the minuscule case this recurrence
relation reduces to
\begin{equation}\label{rec-relb}
\sum_{\nu\in W(\pi )} e^{i\langle \nu ,\xi\rangle}
\mathbf{P}_\lambda (\xi)= \sum_{\begin{subarray}{c} \nu\in W(\pi
)\\ \lambda
+\nu\in\mathcal{P}^+\end{subarray}} \prod_{\begin{subarray}{c} \alpha\in \boldsymbol{R} \\
\langle \nu ,\alpha^\vee\rangle =1 \end{subarray}}
\frac{\sinh\frac{s}{2}(g_{|\alpha |}+\langle \rho_g+\lambda
,\alpha^\vee \rangle )}{\sinh\frac{s}{2}(\langle
\rho_g+\lambda,\alpha^\vee \rangle)}\mathbf{P}_{\lambda
+\nu}(\xi).
\end{equation}
\end{subequations}

\section{Index of Notations}\label{appB}
This Appendix provides a list of notations ordered according to the sections
in which they were first introduced.

\vspace{2ex}
Section \ref{sec2.1}:
$\mathbf{E}$, 
$\langle \cdot ,\cdot \rangle$ ,
$\boldsymbol{R}$, $\boldsymbol{R}^+$, $\mathcal{Q}$, $\mathcal{Q}^+$,
$\mathcal{P}$, $\mathcal{P}^+$, $\alpha^\vee$,
$\mathbf{A}$,  $m_\lambda (\xi)$,  $\xi_w$,
$W$, $W_\lambda$,  $|W_\lambda|$.

\vspace{2ex}
Section \ref{sec2.2}:
$\boldsymbol{R}_0$, $\boldsymbol{R}_1$, $\hat{\Delta}(\xi)$,
$\hat{\mathcal{C}}(\xi)$, $\hat{c}_{|\alpha|}(z)$.

\vspace{2ex}
Section \ref{sec2.3}: $(\cdot ,\cdot)_{\hat{\Delta}}$,
$\text{Vol}(\mathbf{A})$,
$\delta (\xi)$,
$\succeq$, $\geqslant$, $P_\lambda (\xi)$, $a_{\lambda \mu}$.

\vspace{2ex}
Section \ref{sec2.4}: $\chi_\lambda (\xi)$, $(-1)^w$, $\rho$,  $w_\mu$.

\vspace{2ex}
Section \ref{sec3.1}: $\mathcal{H}$, $(\cdot ,\cdot )_{\mathcal{H}}$,
$\hat{\mathcal{H}}$,  $(\cdot ,\cdot )_{\hat{\mathcal{H}}}$,
$\Psi_\lambda (\xi)$, $\mathcal{F}$, 
$\Psi_\lambda^{(0)} (\xi)$, $\mathcal{F}^{(0)}$.

\vspace{2ex}
Section \ref{sec3.2}: $\omega_r$,  $\hat{E}_r(\xi)$, $W(\cdot)$, $L_r$,
$\sigma(L_r)$, $w_0$.

\vspace{2ex}
Section \ref{sec3.3}: $a_{\lambda\mu;r}$,
$\mathcal{P}^+_{\lambda ;r}$, $L_r^{(0)}$.

\vspace{2ex}
Section \ref{sec4.1}: $\mathbf{C}^+$, $m(\lambda)$, $P_\lambda^\infty (\xi)$,
$\|\cdot\|_{\hat{\Delta}}$, $P_\lambda^{m(\lambda)}(\xi)$,
$\Psi_\lambda^\infty(\xi)$, $\hat{S}_w(\xi)$,
$\hat{s}_{|\alpha|}(\langle\alpha, \xi\rangle)$,
$\|\cdot\|_{\hat{\mathcal{H}}}$.

\vspace{2ex}
Section \ref{sec4.2}: $\hat{E}(\xi)$, $L$, $L^{(0)}$, $\sigma (L)$,
$\mathbf{A}_{\text{reg}}$, $\hat{w}_\xi$, $\hat{\mathcal{S}}_L$,
$\|\cdot\|_{\mathcal{H}}$, $\Omega_\pm$, $\mathcal{S}_L$.

\vspace{2ex}
Section \ref{sec5.1}: $\phi^{(0)}(t)$, $\phi_\pm (t)$.

\vspace{2ex}
Section \ref{sec5.2}: $\hat{w}$, $\mathbf{V}_{\text{clas}}$,
$\mathcal{P}^+_{\text{clas}}(t)$, $\phi^{(\text{clas})}(t)$, $\hat{W}$.

\vspace{2ex}
Section \ref{sec5.3}: $\phi_\pm^{(\infty)}$, $P_t^{(\text{clas})}$.

\vspace{2ex}
Section \ref{sec6.1}: $g_{|\alpha|}$,
$(z;q)_\infty$, $\Delta (\lambda)$, $\mathcal{N}_0$,
$\mathcal{C}^\pm(\mathbf{x})$, $\rho_g$, $c^\pm_{|\alpha|}(x)$,
$\mathbf{P}_\lambda(\xi)$.

\vspace{2ex}
Section \ref{sec6.2}: $\pi$, $\alpha_0^\vee$, $\boldsymbol{R}^\vee$,
$\hat{E}_\pi(\xi)$, $L_\pi$, $V_\nu(\mathbf{x})$,
$E_\pi(\mathbf{x})$, $(z:\sinh_s)_m$, $\rho_g^\vee$.

\vspace{2ex}
Section \ref{sec6.3}: $L_\pi^{(0)}$, $n_\pi(\lambda)$, $\alpha_j$,
$r_{\alpha_j}$,
$\mathcal{S}_{L_\pi}$,
$\hat{\mathcal{S}}_{L_\pi}$.

\vspace{2ex}
Section \ref{sec6.4}: $\hat{g}$, $\hat{g}_r$, $(z_1,\ldots,z_k;q)_\infty$,
$g$, $g_r$.

\vspace{2ex}
Section \ref{sec6.5}: $_{s+1}\Phi_s$, $(a;q)_n$, $(a_1,\ldots,a_s;q)_n$.

\bibliographystyle{amsplain}

\end{document}